\documentclass{article}

\usepackage{arxiv}

\usepackage[utf8]{inputenc} % allow utf-8 input
\usepackage[T1]{fontenc}    % use 8-bit T1 fonts
\usepackage{hyperref}       % hyperlinks
\usepackage{url}            % simple URL typesetting
\usepackage{booktabs}       % professional-quality tables
\usepackage{amsfonts}       % blackboard math symbols
\usepackage{nicefrac}       % compact symbols for 1/2, etc.
\usepackage{microtype}      % microtypography
\usepackage{lipsum}

\usepackage[linesnumbered,ruled]{algorithm2e}
\usepackage{tabu}
\usepackage{tabularx}
\usepackage {numprint,siunitx,xcolor}
 \usepackage{makecell,interfaces-makecell}
\usepackage{hyperref}
\hypersetup{
	colorlinks=true,
	linkcolor=blue,
	filecolor=magenta,      
	urlcolor=cyan,
}
\usepackage{enumitem}

\usepackage{subcaption}
\usepackage{rotating}
\usepackage{tikz}
\usepackage{lscape}
\usepackage{amsmath}
\usetikzlibrary{positioning}
\usepackage{array,calc}
\newlength\lena \newlength\lenb \newlength\lenc \newlength\lend
\newcolumntype{P}[1]{>{\centering\arraybackslash}p{#1}} % centered "p" column
\title{A multi-layered blockchain framework for smart mobility data-markets}

\author{
  David Lopez\\
  Laboratory of Innovations in Transportation (LiTrans)\\
  Ryerson University\\
  Toronto, Canada \\
  \texttt{david.lopez@ryerson.ca} \\
   \And
  Bilal Farooq \\
  Laboratory of Innovations in Transportation (LiTrans)\\
  Ryerson University\\
  Toronto, Canada \\
  \texttt{bilal.farooq@ryerson.ca} \\
}

\begin{document}
\maketitle

\begin{abstract}
Blockchain has the potential to render the transaction of information more secure and transparent. Nowadays, transportation data are shared across multiple entities using heterogeneous mediums, from paper collected data to smartphone. Most of this data are stored in central servers that are susceptible to hacks. In some cases shady actors who may have access to such sources, share the mobility data with unwanted third parties. A multi-layered Blockchain framework for Smart Mobility Data-market (BSMD) is presented for addressing the associated privacy, security, management, and scalability challenges. Each participant shares their encrypted data to the blockchain network and can transact information with other participants as long as both parties agree to the transaction rules issued by the owner of the data. Data ownership, transparency, auditability and access control are the core principles of the proposed blockchain for smart mobility data-market. In a case study of real-time mobility data sharing, we demonstrate the performance of BSMD on a 370 nodes blockchain running on heterogeneous and geographically-separated devices communicating on a physical network. We also demonstrate how BSMD ensures the cybersecurity and privacy of individual by safeguarding against spoofing and message interception attacks and providing information access management control.
\end{abstract}


%\end{frontmatter}


\section{Introduction}
\label{sec:introduction}
% Start line numbering here if you want
%\linenumbers

Traditionally, personal mobility data were solicited via small-scale surveys (1-5\% sample) and governments would take the responsibility to secure the personal information before sharing for public use. Nowadays, smartphones, cellphone towers, Wi-Fi hotspots, traffic sensors, among others, can passively solicit detailed mobility data of the urban population. Processing and analyzing passively as well as actively solicited data has the potential to aid governments and researchers to better understand human mobility for designing smarter, demand-driven, reliable and secure transportation systems. To fully exploit the potential of passively solicited large-scale data, privacy and security challenges need to be addressed. Passively solicited data include sensitive personal information like GPS logs or trip and activity habits. Therefore, guarding people's privacy and securing their information from untrusted parties is of utmost importance. 

One of the highest priorities companies are expected to have is the protection of people’s personal information that the companies may collect. Nevertheless, breaches happen whether it is due to poorly designed information systems or hackers finding clever ways of breaching these systems. In recent years, cyber-security breaches have occurred all around the globe and transportation systems are not an exception. In 2015 a group of civic hackers deciphered and exposed the unstandardized bus system location data of Baltimore~\cite{Rector2015}. In 2016 the San Francisco transit was hacked to give free access to commuters for two days~\cite{Stewart2016}. During the same year, information of 57 million Uber customers and drivers were leaked~\cite{Wong2017}.

Data transparency is an important issue in the privacy of individuals. Various services that collect our transportation data fail to clearly explain where or with whom our data are being shared. For example, the Waze app states in the Terms of Service\footnote{\url{https://www.waze.com/en-GB/legal/tos}} that ``...(Waze will) share personal information with companies or organizations connected or affiliated with Waze...''. So all the information collected by Waze could be shared with other organizations associated with Waze. However, the Terms of Service do not make it clear if the user can track with whom their information has been shared.

Another point related to privacy is the access control of the information that is provided by the users. It was recently discovered that Google keeps collecting user location data even if they explicitly deactivate the tracking system in their mobiles~\cite{Nakashima2018}. Another example of disclosing information without the user’s consent is the infamous Facebook-Cambridge Analytica scandal~\cite{Cadwalladr2018}.

The General Data Protection Regulation\footnote{\url{https://eugdpr.org/}} (GDPR) is a step forward for privacy. Although the GDPR is valid only in the European Union, it is still expected to push multinational companies to be more transparent on how they manage people’s private information. In the authors opinion, no matter how many rules and fines governments apply, given the current centralized ways of collecting people’s data, some entities will always find a loophole in legislation or hackers will be able to tamper with these centralized data systems. In addition to laws, individuals have to have total control and ownership of their information. We need to be the guardians and responsible of our own privacy, unfortunately nowadays we have to trust third parties for that.

Distributed ledger technologies like blockchain have the potential to give the people full control of their information, protect individual's personal mobility information and guard their privacy. The technology is difficult to tamper with and transactions are secure as well as transparent to all parties, including the individuals who generated the data. A blockchain is a distributed database, data structure or shared ledger that maintains a list of transaction records, which cannot be altered unless a consensus in the network is reached using an algorithm~\cite{Kim2018}. Some of the most common algorithms in public blockchains are proof-of-work used by Bitcoin and proof-of-stake used by Peercoin, while close blockchains, like Hyperledger, use a byzantine fault tolerance variant. The blockchain is formed by timestamped blocks containing transactions and where each block is permanently linked to a previous block~\cite{Nakamoto2008}. Consensus algorithms together with linked blocks make it very hard to tamper with the blockchain. The network is run by a set of participants so that no single entity controls the flow of information, and nodes can transact with other nodes as long as both nodes agree to some terms, so each node has full control of the assets they possess.

As such, blockchain presents a solution for developing a network for transportation data and associated services, where people own their mobility data, all the transactions are transparent (a public ledger is available to the interested parties in the network), democratic (a consensus must be reached to accept any transaction) and secure (linked blocks make it difficult to tamper with the network). 

In this paper a multi-layered blockchain framework for mobility data transactions is proposed. The main objective is to secure the collected data and to maintain the privacy of the individuals. The rest of the paper is organized as follows: We first introduce the background on how blockchain can solve data management problems and privacy issues in the context of mobility data. We describe the six layer model of the Blockchain framework for smart mobility data transactions. Each layer is described and discussions on different properties of the blockchain are presented. Data shared on the network as well as the rules of participation are discussed. Details of an implementation for the mobility data sharing is outlined and the performance is analyzed. At the end of the paper, a case study and concluding discussion are presented.

%%%%%%%%%%%%%%%%%%%%%%%%%%%%%%%%%%%%%%%%%%%%%%%%%%%%%%%%%%%%%%%%%%%%%%%%%%%%%%%%%%%%%%%
\section{Background}
\label{sec:brackground}
{The distributed data structure of the blockchain was originally developed for the Bitcoin currency ~\cite{Nakamoto2008} as a mechanism to maintain public transaction ledgers. However, in recent years blockchain has gained tremendous attention in other domains e.g. food, pharmacy, real-estate, logistics, etc. This interest is due to blockchain's ability to create secure and private networks where accounting is at the core and no single organization is in control of the transactions as well as the data. Nowadays, distributed ledger technologies are actively developed by industry as well as academia for a range of financial and non-financial applications.}

\subsection{Blockchain in transportation and logistics}
Supply chain management is one of the main transportation applications of blockchain. The stakeholders can track their goods along the complete chain and they do not need to rely on a centralized entity for authenticity of the branded products~\cite{Crosby2016}. In combination with RFID technology~\cite{Androulaki2018,Kim2018}, the blockchain would allow the companies to track products from creation to delivery to the final consumer and will help them to improve their businesses by quickly identifying problems in the chain. 

Blockchain can also be used to tackle transportation supply problems.~\cite{Sharma2017} proposed a blockchain network, where vehicles share their resources (fuel consumption, speed logs, space available, among others) in order to find cheap fuel stations, people for ride-sharing, or to probe good driving behavior in order to get discounts in insurance policies. The Blockchain Mobility Consortium\footnote{\url{https://blockchain-mobility.org/}} wants to share and monetize the driver's information to improve network performance and to make money while driving. Applications like Arcade City\footnote{\url{https://arcade.city/}} are proposing to share their trips in a shared mobility service, but without third party involvement in the transaction. Shared mobility can exploit the use of blockchain to connect drivers and riders with no third party intermediaries. However as~\cite{Stocker2016} pointed out, some issues like regulatory uncertainty, liability issues and network optimization need to be address before fully implementing blockchain for shared mobility. The blockchain B\textsuperscript{2}ITS is a conceptual model which can be used as a network for Parallel Transportation Management and Control System~\cite{Yuan2016}.

\subsection{Privacy and cybersecurity}
Maintaining individual's privacy is currently one of the key challenges faced by various industries and researchers. Almost every part of our lives is stored on servers owned by various companies. Previously, techniques like hashing function have been used to anonymize user data in transportation \cite{beaulieu2016large}. Public and private key encryption techniques have been used for data and communication security~\cite{Farooq2015}. However, researchers and technologists have found that blockchain can be a potential solution to the privacy problem by decentralizing information and making the individuals the sole owners and controllers of their information. Blockchain can be used to securely share private information in: medical networks~\cite{Yue2016}, IoT networks~\cite{Dorri2017}, smart grids~\cite{ZhumabekulyAitzhan2016} and data provenance in cloud computing~\cite{Liang2017}.

To the authors knowledge, in the literature there is no record of a generalized multi-layered blockchain framework for smart mobility data transactions that can guard the privacy of individuals and protect against hacking. The framework presented in this work can be a solution to the privacy and security challenges of sharing actively as well as passively solicited large-scale smart mobility data.

%%%%%%%%%%%%%%%%%%%%%%%%%%%%%%%%%%%%%%%%%%%%%%%%%%%%%%%%%%%%%%%%%%%%%%%%%%%%%%%%%%%%%%%
\section{Conceptual framework}
\label{sec:framework}
The level of permission is the first step in the creation of a blockchain, e.g., in the Bitcoin blockchain anyone can participate and all the transactions are publicly available. Essentially there are four types of permissions~\cite{Olnes2017}:
\begin{enumerate}
	\item \emph{Public closed}: Anyone can do the transactions and have access to the ledger. Only a restricted set of participants can be involved in the consensus mechanism.
	\item \emph{Public open}: Anyone can do the transactions, have access to the ledger and can participate in consensus mechanisms. 
	\item \emph{Private closed}: Restricted access to transactions, have access to the ledger and the consensus mechanisms. Only the owner determines who can participate.
	\item \emph{Private open}: Restriction on access and who can transact. All participants can be involved in the consensus mechanism.
\end{enumerate}

The main disadvantage of \emph{open} blockchains is the amount of energy to reach consensus necessary to build trust between all parties in the network, for example the estimated energy consumption of the Bitcoin is 100MW~\cite{Harald2017}. In the authors' opinion it will be better to consider an eco-friendlier path and to opt for a network with less consumption of energy. We propose a \emph{close} blockchain as this type of network consumes considerably less energy than an \emph{open} counterpart. 

The decision between \emph{private} or \emph{public} has to be taken in terms of access to the ledger and participation. In \emph{private} only the owner determines who can participate, giving such power to one entity or entities may lead to an undemocratic process and can hurt people's trust. The \emph{public} blockchain may fit better in the proposed framework as the participation is open to the public. However, in this type of networks the ledger is also public, so in order to protect the privacy of the individuals the personal information is not stored in the ledger. Nevertheless, we believe that choosing the right type of blockchain, should be a decision taken by all parties involved. People, society, government bodies, and concerned companies need to discuss this in depth.

\begin{figure}
	\centering
	\centerline{\includegraphics[scale = .16]{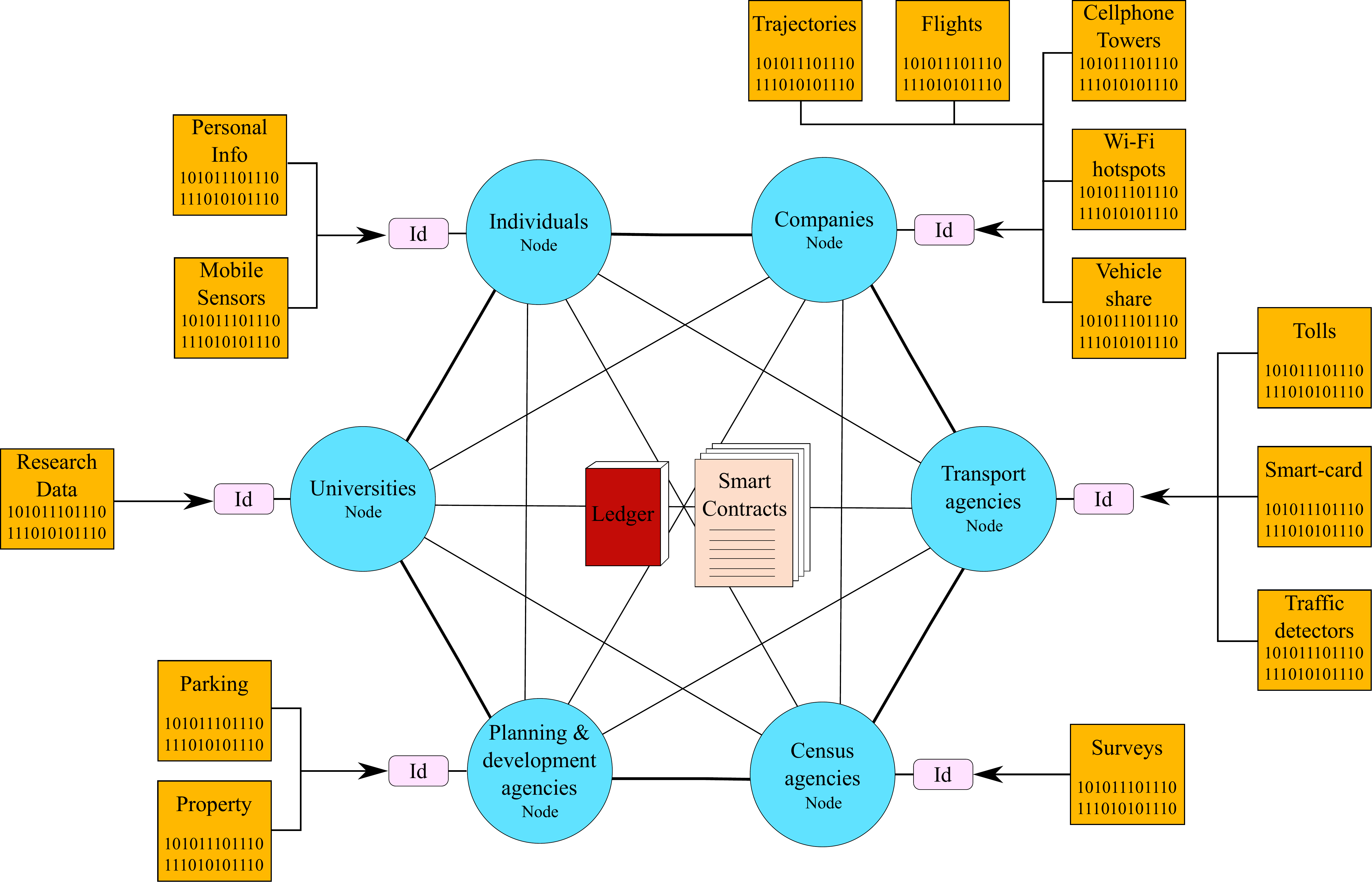}}
	\caption{Blockchain network and data collectors}
	\label{fig:diagramBlock}
\end{figure}

Figure~\ref{fig:diagramBlock} shows the general framework of a \emph{public closed} Blockchain for Smart Mobility Data-market (BSMD) composed of nodes: \emph{Individuals}, \emph{Companies}, \emph{Universities} and \emph{Government} (transport, census, planing and development agencies). The nodes collect their own data and store it in \textit{identifications}. Each node is the sole owner of their data and can share their information by showing other nodes their \emph{identifications} or parts of it.   

Nodes in BSMD are divided into \emph{passive} nodes and \emph{active} nodes. \emph{Passive} nodes may read or host copies of the ledger. This type of node is suitable for individuals or small businesses who want to participate and take advantage of the network, but do not have the resources for running nodes for extended periods of time. \emph{Active} nodes can write blocks and store updated versions of the ledger for other nodes to connect. This type of node is suitable for governments, universities or companies who have the resources for these tasks. In the blockchain there are \emph{smart contracts} available that the nodes need to sign before any transaction of information is conducted. 

{There are no differences between the \emph{active} and \emph{passive} nodes in terms of their ability to do the transaction of information on BSMD. An individual person can be an \emph{active} node, if they wish to. However, given that the BSMD is \emph{closed} they would need to acquire permission from the current nodes to become \emph{active}. Furthermore, it is expected from them to have a strong computing power available.}

\begin{figure}
	\centering
	\centerline{\includegraphics[scale = .27]{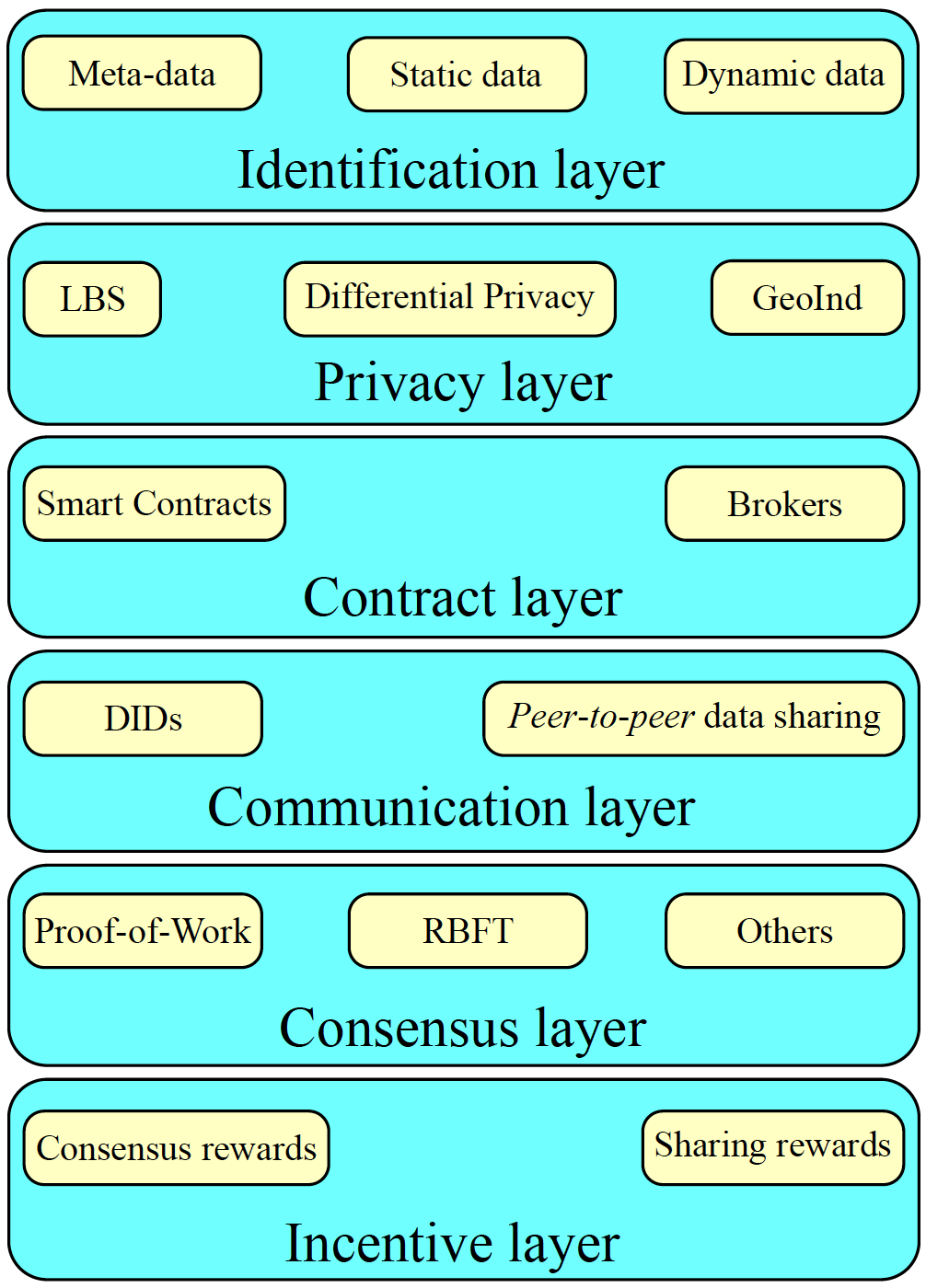}}
	\caption{Multi-layered Blockchain model for smart mobility data-market.}
	\label{fig:layerBlock}
\end{figure}

{Blockchain frameworks are often described in layers~\cite{Biswas2016, Glaser2017, Yuan2016}, inspired from the classic Open System Interconnection (OSI) model~\cite{day1983osi}. Hence in Figure~\ref{fig:layerBlock} we present a six layered model for BSMD}. The Identification layer is composed of mobility and other information that the nodes own. The Privacy layer is the differential privacy model for accessing Location Based Services. In the Contract layer are the set of \emph{smart contracts} and the \emph{brokers} who facilitate data transactions between nodes. The Communication layer contains the Decentralized Identifiers~\cite{Reed2018} of the nodes whose serve as endpoints to establish \emph{peer-to-peer} connections. The Consensus layer contains the consensus algorithms in which the \emph{active} nodes agree to write transactions in the ledger. Finally, in the Incentive layer are the rewards the \emph{active} nodes receive for participating in consensus and the reward nodes receive for sharing (selling) their information. In the following subsections each layer is discussed in greater details.

%%%%%%%%%%%%%%%%%%%%%%%%%%%%%%%%%%%%%%%%%%%%%%%%%%%%%%%%%%%%%%%%%%%%%%%%%%%%%%%%%%%%%%%
\subsection{Identification layer}
\label{sec:identificaton}

Mobility data is constantly generated by different nodes. There are several companies, municipalities, and individuals producing transportation information which is valuable to governments, researchers and people. For example, telecommunication companies generate data that can be used for transportation modeling, the logs of available mobile devices registered by cellphone towers or Wi-Fi hotspots can be used to monitor traffic~\cite{Farooq2015} or to capture the individual's daily activity patterns~\cite{Phithakkitnukoon2010}. The companies can also take advantage of the blockchain to find customers or use the data generated by Government, Universities or other Companies to improve their business. It is worth noting that according to the BSMD framework companies are in control of their data so they can decide to what extent they want to share information.

One of the responsibilities of the government is to collect data in order to model, manage and improve transportation networks. Information on tolls, smartcards, traffic detectors, surveys, parking and properties can be use to find new ways of shaping our mobility~\cite{Toledo2017,Alsger2018}.

Universities often need to collect particular data not collected by government or companies. This data is often targeted to specific purposes like the state-preferences survey on the willingness to buy an autonomous vehicle~\cite{farooq2018virtual}. 

The biggest data source of transportation systems are the individuals and their smartphones. Everyday individuals generate huge amounts of data which is passively or actively solicited by companies or governments. Nowadays, it is possible to infer different aspects of the travel behavior of the users just by processing the GPS information gathered using their smartphones \cite{Yazdizadeh2018}. 

{An important principle of BSMD is that nodes are in control and can track their data}. All transportation data are contained in files called \emph{Identification} and they are stored in personal cloud services or in personal devices. The advantage of the cloud services is that data generated by mobiles is updated in real-time and is accessible from any device connected to the web.

Each \emph{Identification} is composed of metadata, static data and dynamic data. The metadata contains the information relative to the characteristics of the node, but without disclosing any personal information. This information is publicly available and is the gateway to find nodes in the network. By looking into the metadata other nodes can see the blockchain address for making connection request and if the node they are trying to reach is an individual, a government agency, a private company or an academic institution. 

{Nodes may opt to request that their identity key, issued by a trusted node (e.g. government, reputable non-profit agency, university, etc) is exposed in the metadata. This key will give data generator nodes some level of confidence that the node they are dealing with is an actual person/agency/entity with verifiable credentials. Issuing identity keys would work the same way as in the current situation, where every government issues IDs to individuals, companies, etc. residing/working in a country. For an individual to acquire their identity key, it would be necessary to show the trusted nodes (e.g. government) their birth certificates, passports or other government issued papers. Identity keys of the individuals can be used to show that the data owner is a verifiable-person. It can also be used to verify other attributes, e.g. age, gender, location etc. Companies/agencies/organizations may need to show paper work from the public registry in order to get the identity key. They can prove their sector (educational, commercial, government), address and contact information with their keys. Individuals and companies/agencies/organizations need to show a trusted node, actual identifications that are supposedly difficult to tamper with, e.g, a passport or public registry records. If nodes are trying to fake their identity in the BSMD, they will have to first fake their real identification papers. The identity keys have been used to verify claims with the issuer in other studies~\cite{SovrinFoundation2018}. For illustration purposes, suppose that nodes A and B want to share some information. First, node A would ask B to prove that it is an actual person. At the time of joining, node B would have already been issued their identity key by showing their passport to the government. Node B sends their key to A and A uses this key to verify in the BSMD that B is an actual person. At the same time, A would not know the personal attributes e.g. gender or age of Node B, if they are not shared in the metadata. Keys do not contain personal information, but can be used to prove claims without disclosing personal data. To make it further secure, the issuing node can keep the identity keys for the verified-nodes dynamic.}

\begin{figure}[!h]
	\centering
	\centerline{\includegraphics[scale = .350]{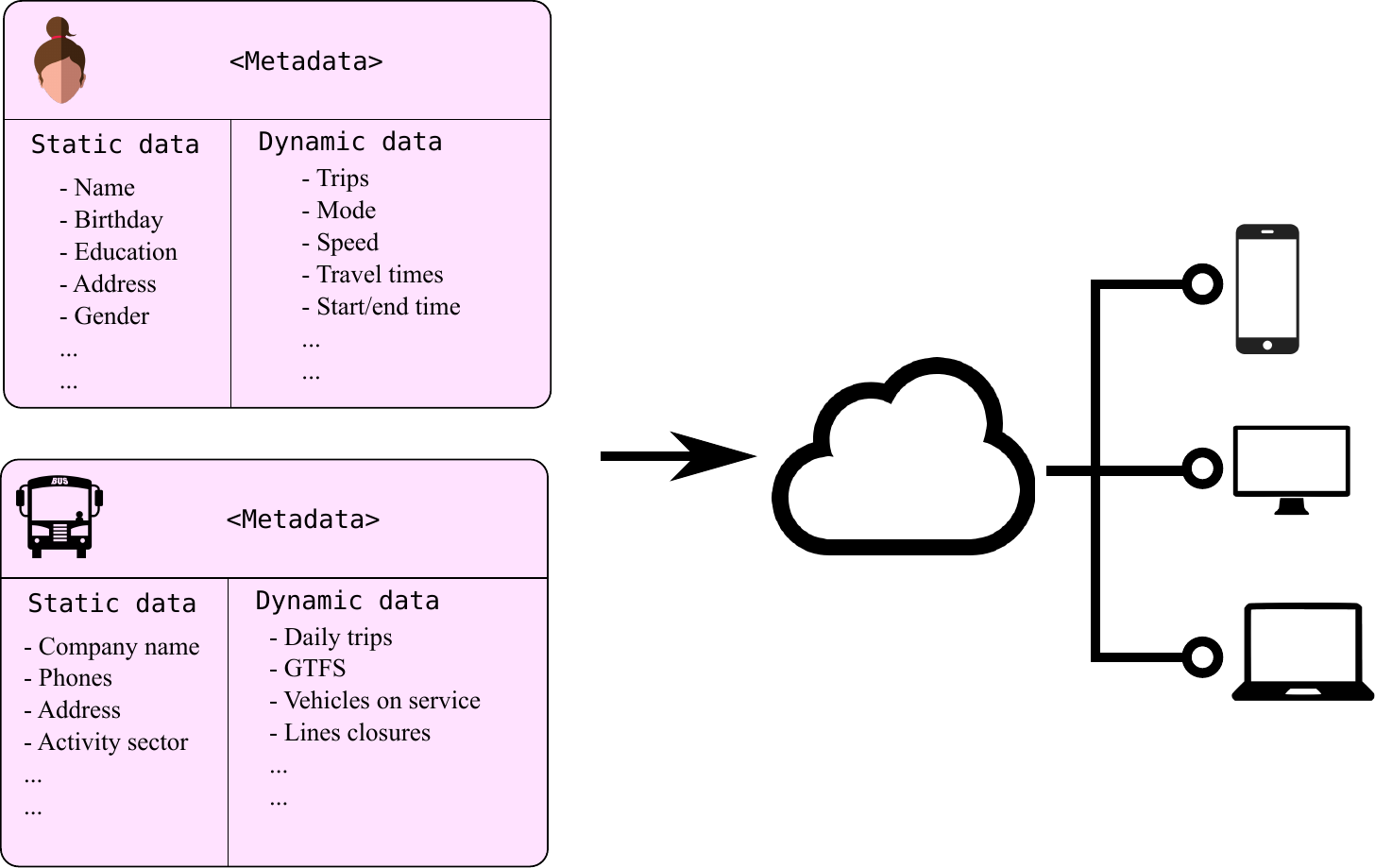}}
	\caption{\emph{Identification} schema}
	\label{fig:identification}
\end{figure}

The static data in \emph{Identification} layer contain information that does not change for long periods of time. The static information of an individual can be their name, birthday, gender, etc. While the static information of a government agency can be name, sector, board members, etc. The dynamic data contains information that is generated continuously. The dynamic information of an individual can be: daily trips, speed, modes used, origins-destinations, among others. The dynamic information of a transit agency are their daily demand and supply, real-time information on arrivals, vehicles in services, line closures, etc. Nodes need consent from the owner to access the static and the dynamic data. Figure~\ref{fig:identification} shows an \emph{identification} file of an individual and a transit agency. 

%%%%%%%%%%%%%%%%%%%%%%%%%%%%%%%%%%%%%%%%%%%%%%%%%%%%%%%%%%%%%%%%%%%%%%%%%%%%%%%%%%%%%%%
\subsection{Privacy layer}
\label{sec:privacy}

There is no doubt that Location Based Services (LBS) have made our life easier as we have instant access to information about our surroundings. However, when we access those services we are fully disclosing our location to receive the information we are looking for. For example, if users want to know the arrival times of bus lines near their location they will need to disclose their actual location in order to receive the information they want. BSMD will protect users from external agents trying to steal their information as all communications between nodes in the blockchain are \emph{peer-to-peer} and the records in the ledger do not contain information that can be used to track the location of a specific user. However, the provider of the LBS will know the actual location of the user. No matter if the node providing the LBS is honest or not, they still know the actual location of the user and this could be perceived as a privacy invasion.

It is possible to access LBS without disclosing the user's exact location. One popular technique is $k$-anonymity~\cite{Sweeney2002} which consist of hiding the real location in a set of similar, but fake locations. Thus, making it hard to identify the real data. This technique has been applied in LBS through the use of: \emph{dummy locations}~\cite{Lu2008} where the dummy locations along with the real locations are sent together to the LBS; \emph{cloaks}~\cite{Zhao2018} where a region containing the real location is sent to the LBS; and \emph{geomasking}~\cite{Zhang2017} where the real locations is randomly displaced outside of an inner circle but contained in an outer circle, i.e., the real location is displaced inside a ``donut''. 

\begin{figure}[!h]
	\centering
	\centerline{\includegraphics[scale = .40]{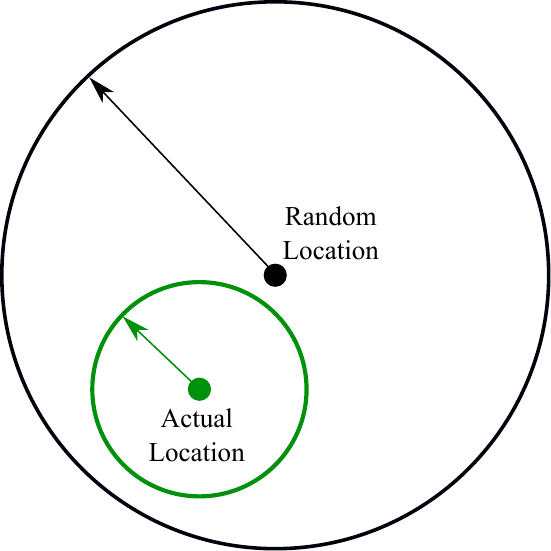}}
	\caption{Search radii in \emph{GeoInd}}
	\label{fig:geoind}
\end{figure}

Another model which is gaining momentum in recent years is Differential Privacy~\cite{Dwork2008}. Apple has started using this model to anonymize mobile usage while typing\footnote{\url{https://www.apple.com/ca/privacy/}}. In Differential Privacy the probability of a query returning a value $v$ when applied to a database $D$ is similar when compared to the probability to query the same value in an adjacent database $D'$ differing by only one observation. When the operations are performed on $D$ the outcome of these operations will be close enough to the outcome of the operations performed on $D'$. In the context of LBS a Differential Privacy model called \emph{Geo-indistinguishability} (\emph{GeoInd}) can be used~\cite{Andres2013}. When a user query LBS using \emph{GeoInd} instead of searching from the actual location they will use a random nearby location such that it is possible to filter the results from the fake location to get information they are looking for. The idea is that from the random location the larger search radius will contain the smaller search radius of the real location. In Figure~\ref{fig:geoind} the actual location and search radius are shown in green and the random nearby location and search radius are shown in black. The user sends the random location (black dot) to the LBS to get the result contained in the black circle. Then, the user will filter the results of the black circle to get the results contained in the green circle. In this manner, the actual location of the user will never be sent to the LBS.  

%%Bilal: I found other methods (see bellow) which use Differential Privacy in LBS however in my opinion referencing those methods is out of the scope, as this paper is not intended to be a state-of-the-art in Differential Privacy and LBS
%% -https://dl.acm.org/citation.cfm?id=3137804
%% -https://ieeexplore.ieee.org/abstract/document/8413135/

\cite{Oya2017} reported that for maintaining a high level of privacy in \emph{GeoInd} the utility must be sacrificed. Thus, the information retrieved from LBS using \emph{GeoInd} may not be useful unless users sacrifice some degree of privacy. On the other hand~\cite{Badu-Marfo2018} reported that although $k$-anonymity using \emph{geomask} outperformed \emph{GeoInd} in privacy protection, the latter shows promising results for data utility if properly configured. 

Nowadays, accessing LBS is part of people's day-to-day life. Whether it is for looking for restaurants or searching bus arrivals, but at the same time users prefer not to disclose their location. In BSDM, we propose a hybrid approach, where users can select the level of privacy. $k$-anonymity using \emph{geomask} is used for high levels of privacy and low levels of utility while \emph{GeoInd} is used for low levels of privacy and high levels of utility. In situations where the exact location is not important for getting accurate response, \emph{geomask} is used to protect the privacy, e.g. getting information of subway network. In situations where the exact location is important for getting accurate response, \emph{GeoInd} is used, e.g. walking distance to nearest subway station. Also, the LBS provider can reward users for their information if more relaxed parameters for privacy are set, but in the end the user is in control of their privacy and not the LBS provider.

%%%%%%%%%%%%%%%%%%%%%%%%%%%%%%%%%%%%%%%%%%%%%%%%%%%%%%%%%%%%%%%%%%%%%%%%%%%%%%%%%%%%%%%
\subsection{Contract layer}
\label{sec:contracts}

The contract layer is composed by \emph{Smart Contracts} and data \emph{Brokers}. A \emph{smart contract} is a set of promises, specified in digital form, including protocols within which the parties perform on these promises~\cite{Szabo1996}. A \emph{smart contract} is a script which defines the set of assets available to transfer and the type of transactions permitted. All \emph{smart contracts} are stored in the blockchain and have a unique address. They act as independent actors whose objective is to transact assets given a certain set of rules that involved parties agreed upon. Once all the parties agree on the terms of the contract, they cryptographically sign the \emph{smart contracts} and start sharing the selected data via a \emph{peer-to-peer} connection. 

{In BSMD, \emph{smart contracts} are used to self-enforce fair trade between the nodes and to automatically solve various disputes. When nodes join the BSMD, they select their terms for the data they are willing to share and such terms cannot be altered by other nodes. For instance, if a node considers the information they are willing to share is not valuable for them or might not affect their privacy, they might give it away for free. On the other hand, if the node considers that they are sharing some personal information with a high value, they might ask for a reward in exchange and share their information using differential privacy. Information acquirer can also set their own terms for the transactions e.g. maximum payments, type and value of the service to be provided, and the accuracy level they want from the collected information.}

{Before a transaction is executed, the terms of sharing from all the parties involved are considered to automatically select the appropriate \emph{smart contract} that can handle the transaction. If the terms of the parties conflict the transaction is either not performed or parties are given a chance to negotiate and update their terms. The advantage of \emph{smart contracts} is that minimal human interaction is need. Nodes just need to setup their terms once and maybe renegotiate a contract if the new terms are in their best interest. Predefined terms can be given to individual users when entering the BSMD or users may opt to use advanced options and create an \emph{ad-hoc} set of terms. After nodes select their terms, transaction procedures are transparent to the nodes and self enforced by \emph{smart contracts}}.

{In general, the terms of the owner of information should be composed of:} 
\begin{enumerate}
	
	\item \emph{Service requested}: Information/services the user wants in exchange for the data. 
	\item \emph{Monetary reward}: Payment for the shared data. 
	\item \emph{Level of privacy}: To what extent the owner is willing to disclose. Owner may opt to share raw or aggregated information, e.g. user may disclose their age or range of age.
	\item \emph{Temporality}: For how long the owner is willing to share his data. For example, share information during April or just one time.
	\item \emph{Extended permissions}: Owner can select the level of redistribution. For example, owner may forbid or give permissions to redistribute their data for marketing purposes.
	\item \emph{Identity key}: Nodes may require to validate the identity key of the requester before starting the sharing process.
\end{enumerate}

{While the contract terms for an acquirer/requester should be composed of:}
\begin{enumerate}
	
	\item \emph{Service provided}: Information the requester will share in exchange for the data. 
	\item \emph{Monetary reward}: Payment value to the owner for their data.
	\item \emph{Accuracy of information}: Level of detail of the requested data. For example, a company may solicit exact location of the owner.
	\item \emph{Temporality}: For how long the requester wants to get the data of the owner. 
	\item \emph{Extended permissions}: With whom it will share/sell the owner data
	\item \emph{Identity key}: Nodes may require to validate the identity key of the owner before starting the sharing process.
\end{enumerate}

{\emph{Smart contracts} would reside on the blockchain and are automatically activated before a transaction. A typical \emph{smart contract} in the BSMD is composed of the terms of both parties and \emph{conditional-statements} deciding if the transactions can be perform. The \emph{smart contract} shown in Algorithm~\ref{alg:smart} takes as input the terms of the owner and requester nodes. If both terms are in line with each other the nodes digitally sign the contract and a \emph{peer-to-peer} connection between the nodes is opened to complete the transaction. If at least one term (from the owner or requester) is not fulfilled, the connection is rejected.}

\begin{algorithm}
	\footnotesize
	\SetAlgoLined
	($s_r$)= \{get near by restaurants ($s_o$), no reward ($r_o$), differential privacy is use and age range ($p_o$), one time share ($t_o$), identity key required ($id_o$)\}\\
	Requester terms = \{send near by restaurants ($s_r$), no reward ($r_r$), low geographic accuracy ($a_r$), one time share ($t_r$), identity key required ($id_r$)\}\\
	\SetKwProg{Fn}{Function}{}{end}
	\Fn{ShareInfo(Owner terms, Requester terms)}{
		\eIf{$s_o$ = $s_r$}{
			\eIf{$r_o \leq s_r$}{
				\eIf{$p_o \subset a_r$}{
					\eIf{$t_o \subset t_r$}{
						\eIf{$id_o$ pass verification AND $id_r$ pass verification}{
							continue connection;
						}
						{
							refuse connection;
						}
					}{
						refuse connection;
					}
				}{
					refuse connection;
				}
			}{
				refuse connection;
			}
		}{
			refuse connection;
		}
	}
	\caption{\emph{Smart contract} Information transaction}
	\label{alg:smart}
\end{algorithm}

\emph{Brokers} are nodes in the network that arrange transactions between nodes for selling or buying transport information. The \emph{Brokers} can be associated to companies nodes or to individual nodes (see Figure~\ref{fig:diagramBlock}). They do not participate in data transfers and do not have access to private data of their clients. Their job is to find pairs of nodes that benefit from the exchange of information. To reduce possible scams, \emph{brokers} need identity keys from trusted nodes {(e.g. government, university, non-profit organization)}, so when a \emph{broker} contacts a node, before making a connection the node may will a trusted node if the \emph{broker} has an identity key, otherwise the connection is rejected. {When \emph{brokers} are involved in the transaction between two nodes they can set in the \emph{smart contract} a fee that they would collect for every transaction. Later in Section~\ref{sec:implementation} an example is given that describes the complete transaction process when one of the nodes request the services of a \emph{broker}.}

\begin{figure}
	\centering
	\centerline{\includegraphics[scale = .5]{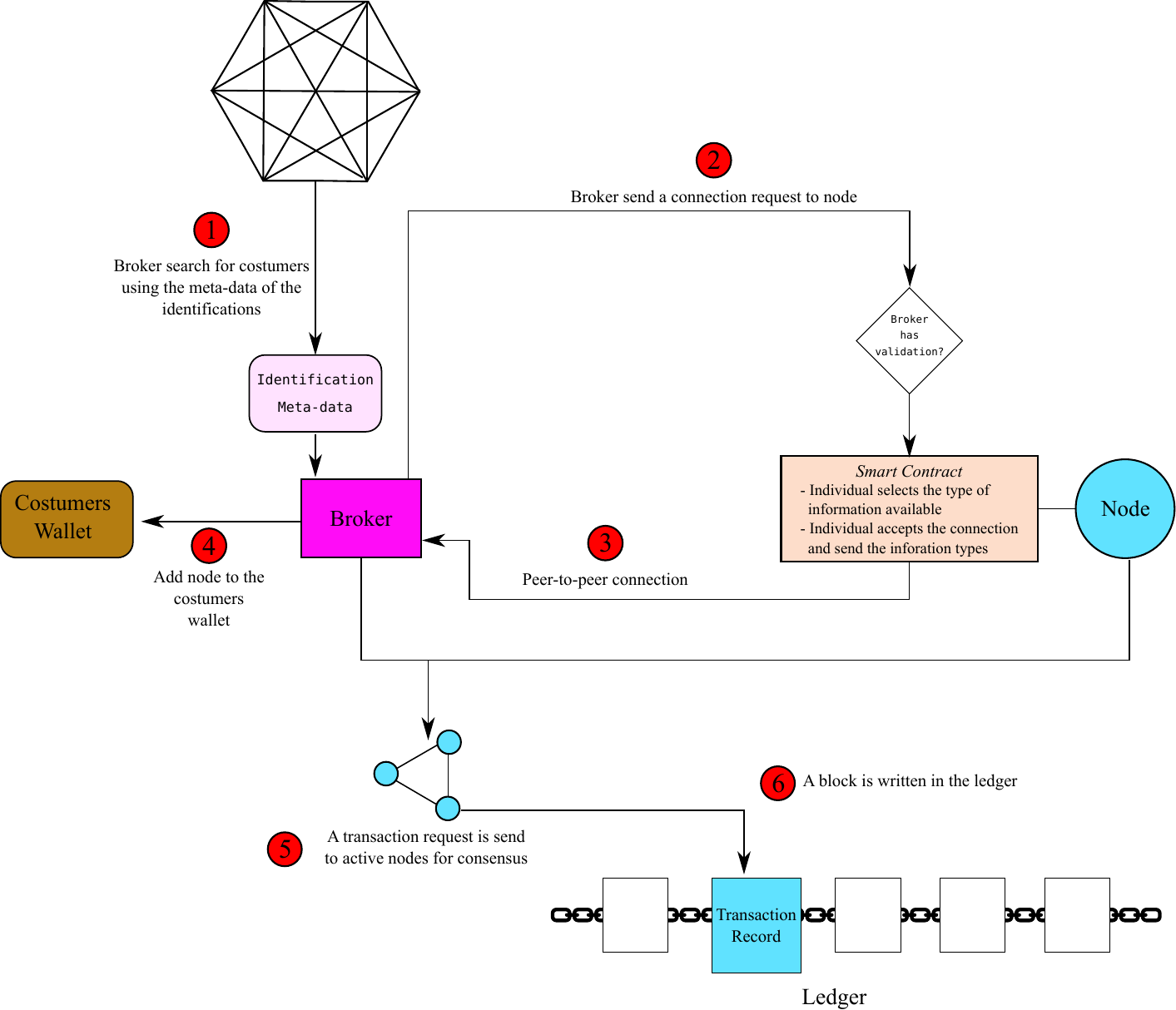}}
	\caption{\emph{Broker} searching and getting costumers}
	\label{fig:broker}
\end{figure}

\emph{Brokers} look for costumers by exploring the meta-data of the nodes and then ask them the type of information they are willing to share or the information they may need from other nodes. Figure~\ref{fig:broker} shows the \emph{broker} process for searching and getting costumers. In the step $1$ the \emph{broker} searches in the blockchain for costumers using the  identification metadata. In step $2$ it will try to connect with the nodes using the metadata address, but before a communication can be established the node will check if the \emph{broker} has validation form a trusted node. Both parties need to sign a \emph{smart contract} to establish a connection. In step $3$ the \emph{broker} and the node communicate and the \emph{broker} saves their costumer in a wallet. In step $5$ both parties communicate the transaction to \emph{active} nodes for consensus (see Section~\ref{sec:consensus}). Finally in step $6$ an \emph{active} node writes the transaction in the ledger.

%%%%%%%%%%%%%%%%%%%%%%%%%%%%%%%%%%%%%%%%%%%%%%%%%%%%%%%%%%%%%%%%%%%%%%%%%%%%%%%%%%%%%%%
\subsection{Communication layer}
\label{sec:Communication}
Nodes communicate with each other using \emph{Decentralized Identifiers (DID)} which are fully under their control, independent from any centralized registry, identity provider, or certificate authority~\cite{Reed2018}. \emph{DIDs} are URLs to communicate with other nodes with the advantage that any node can create their own DID without the permission of a central authority.

The address in the \emph{identification} metadata is a Decentralized Identifier called \emph{DID$_p$} which resolved to a document that contains: (a) the service endpoint for requesting connections and (b) the characteristics of the node but without disclosing any personal information. As all \emph{DID$_p$} are public this cannot be used for sharing transport data.

Once two nodes accept to share information a new \emph{DID$_i$} is created by each node for opening a communication channel and sharing the information. So a single node will have multiple \emph{DID}s to communicate with the nodes. Having multiple $DID$s for communication will make it harder to intercept the information while it is being transferred. Instead of tampering one communication channel the attacker will need to tamper multiple communication channels. An advantage of the \emph{DID}s is that its owners can revoke access as they please~\cite{Reed2018}. Hence, at any given point a node can remove their \emph{DID} and shutdown the communication. Revocability of access to personal data is one of the key aspects in data privacy, which unfortunately in today's world is not that simple. On-line search shows several sites with detailed instructions for opting out from different pages. With the use of DID opting out is as simple as deleting a file on your computer. {$DID$s are the gates for sharing data via \emph{peer-to-peer} connections where the information is transferred using an asymmetric encryption.}

{In BSMD the ledger is public, hence everyone can query its contents. If the information being transacted is of public interest, like the location of transit vehicles during service hours, then all the public information (without duplication) along with the \emph{DID}s are written in the ledger. When the transacted information is private, only the \emph{DID}s of both nodes along with the type of information being transacted is written in the ledger}. Given that a node will have one unique DID per transaction, it is difficult for an attacker to correlate \emph{DID}s in the ledger to track single nodes.

%%%%%%%%%%%%%%%%%%%%%%%%%%%%%%%%%%%%%%%%%%%%%%%%%%%%%%%%%%%%%%%%%%%%%%%%%%%%%%%%%%%%%%%
\subsection{Incentive layer}
\label{sec:incentive}
Incentives are economic rewards for the participants of the network. They have been used in different blockchain projects~\cite{Shea2017,ODEM2018,Emery2018} as a way to motivate the participants to stay and maintain the network. For example, Blockchains for Cryptocurrency reward participants for validating transactions in the network. In BSMD one reward is given to \emph{active} nodes for participating in the consensus mechanisms and write blocks in the ledger. Incentives for hosting the network and participating in consensus mechanisms motivate the nodes to maintain the network and secure the transactions.

Other rewards are given for sharing (selling) information. Any node can put a price on their information or can make an offer to other nodes to access their information. The main business of \emph{brokers} is in this layer, as they are supposed to know how the market is moving in order to get their costumers the best deal for selling or buying transportation data. For every transaction of a costumer the \emph{broker} receives the percentage specified in the \emph{smart contract}.

{In BSMD, users own their information so they can decide whether to sell/share it or not. This rule contradicts the current practices where companies permanently own our information and in exchange they provide us a service. Nowadays, the mobility service (e.g. dock-less scooters, routing applications, LBS, among others) providers are not only interested in the rental of their equipment, but also in the collection of disaggregated mobility patterns of their users. Such information can be used for advertising as well as it can be sold to third parties. So a question arises: If users have control over their information would companies still be able to make profits, while offering rewards in exchange for personal mobility data? In~\ref{sec:appedixgame} we analyzed this question using game theory and discuss how and under what conditions companies can still make profits when users own and sell their information.}
%%%%%%%%%%%%%%%%%%%%%%%%%%%%%%%%%%%%%%%%%%%%%%%%%%%%%%%%%%%%%%%%%%%%%%%%%%%%%%%%%%%%%%%
\subsection{Consensus layer}
\label{sec:consensus}
When a transaction is performed between two nodes a block is written in the ledger. However, before writing a block the nodes in the network need to reach a consensus. Selecting the most adequate consensus algorithms will depend on the level of security, the energy consumption, and the trustiness of the nodes, a comparative analysis of some consensus algorithms can be found in~\cite{Zheng2017}. 

In general the process to reach consensus is the following. When two nodes share information a transaction state is sent to the nodes in the network. Given that state some or all nodes performs a computation and share their response to the network. Using the responses from the nodes a consensus is reached and an \emph{active} node writes a block in the ledger. This new block contains the DID associated with the nodes that are sharing information, the type of information being transacted and if necessary the DID of the \emph{broker} who facilitated such transaction. In the BSDM only \emph{active} nodes can write blocks in the ledger as they are the only ones with the infrastructure necessary to make intensive computations and to host complete copies of the ledger.

Next, we outline some common consensus algorithms and then we discus the selection of an algorithm. In Proof-of-work (PoW) one node solves a computer intensive puzzle based on the transaction and publishes their response in the network where other nodes can verify the result. If the result is verified the node can write a block in the ledger~\cite{Nakamoto2008}. In Proof-of-Stake (PoS) a lottery is drawn between all the nodes and the prize for the winner is to write a block in the ledger. The number of tickets each node receives is proportional to the stake they have in the network, so the more stake a node has the better their chances are of winning the lottery~\cite{Zheng2017}. Practical Byzantine Fault Tolerance (pBFT) and Tendermint are byzantine algorithms, where nodes are voted through different stages and at the end a node is selected to write a block in the ledger, usually nodes with better reputation (nodes that never write faulty blocks) will have more votes.

Table~\ref{tab:consensus} shows a comparison of the PoW, PoS, pBFT and Tendermint consensus algorithms~\cite{Zheng2017}. The level of permission row indicates the openness to participate in the consensus mechanism. The energy saving row indicates the resources necessary to reach consensus. For instance the PoW algorithm consumes a considerable amount of resources while the pBFT and Tedermint consume less resources compared to the others. The tolerated power of adversary row indicates the amount of control an attacker would need to forge transactions. PoW requires that a single node have over the $25\%$ of the computing power of the network\footnote{\cite{Eyal2018} showed that by using \emph{selfish-mining}, it is possible for a set of nodes to control the Bitcoin network}, in PoS a single node would need more than $51\%$ of the stakes, pBFT and Tendermint need that more than $33.3\%$ of the nodes send incorrect messages.

\tabulinesep=1.5mm
\begin{table}
	\small
	\caption{Consensus algorithm characteristics~\cite{Zheng2017}}
	\label{tab:consensus}
	\small
	\begin{tabu} to \linewidth  {| X[l,m] | X[l,m] | X[l,m] | X[l,m] | X[l,m] |}
		%%\begin{tabular}{m{6em} m{5em} m{5em} m{5em} m{5em} m{5em} }
		\hline
		\multicolumn{1}{|c|}{\textbf{Property}} & \multicolumn{1}{c|}{\textbf{PoW}} & 
		\multicolumn{1}{c|}{\textbf{PoS}} & \multicolumn{1}{c|}{\textbf{pBFT}} & \multicolumn{1}{c|}{\textbf{Tendermint}} \\
		\hline
		Level of permission & open & open    & permissioned & permissioned      \\ 
		\hline
		Energy Saving            & no   & partial & yes          & yes    \\ 
		\hline
		Tolerated power of adversary  & $<25\%$ computing power & $<51\%$ stakes & $<33\%$ faulty replicas & $<33\%$ voting power                    \\ 
		\hline
	\end{tabu}
	
\end{table}

{BSMD is a \emph{public closed} blockchain which means that the nodes will need permissions for participating in consensus mechanisms. We propose the creation of a consortium formed by government, universities, transportation companies and non-profit organizations, who will be in charge of managing the submission process for becoming and \emph{active} node.}  

To participate in consensus mechanism a node will have restrictions that depend on the implemented algorithm for consensus. For instance, although the PoW and PoS algorithms are open, in BSMD if a node wants to participate in consensus their capacity will be restricted and by law they will not be able to control more than $25\%$ of computing power in PoW or $51\%$ of the stake in PoS. Also any kind of alliances will be prohibited, if this means that the new alliance will lead to a control over network. Analogous restriction can be set for the pBFT and Tendermint algorithms.

Unlike cryptocurrencies where all nodes are anonymous, in BSMD the nodes that participate in consensus are not. This will guarantee that nodes engaging in obscure practices during the consensus are easily identified and fines can be applied to those guilty of such practices.

In addition to knowing who can participate in consensus, another point that has to be taken into account is the energy consumption and the scalability of the network. BSMD could be grown to a national scale, so it is desirable that the energy necessary to run the network does not impact the environment or number of transaction per second, which will be considerable. PoW and PoS suffer from energy issues (see Table~\ref{tab:consensus}). However, PoS energy problems are not as bad as in PoW. Scalability issues in blockchain an area of ongoing research. For instance, PoW only permits 7 transaction per second~\cite{Vukolic2016}, while variants of Byzantine Fault Tolerant (BFT) algorithms can handle up to 3,500 transaction per second~\cite{Androulaki2018}. 
{As BSMD is a \emph{public closed} blockchain, Byzantine Fault Tolerant (BFT) algorithm is a good option for consensus. Location Base Services (LBS) will consume most of the transactions in the BSMD, because these services tracks people in real-time to provide services like traffic alerts or turn-by-turn navigation. Even if in BSMD, $75\%$ of the users queries are LBS, it can currently handle $76,000$ users in real-time and with at least 90\% throughput (see~\ref{sec:appedix} for the mathematical analysis). Scaling up of the consensus mechanism in the blockchain is an active area of research. In near future, BSMD will be able to take advantage of these developments to manage even larger number of users in real-time.}

Nowadays, the number of transaction per second is a counterweight for implementing blockchains on metropolis. However, some promising results are coming to light~\cite{Zhang2018,Sousa2017} that may solve the problem of scalability in coming years.

%%%%%%%%%%%%%%%%%%%%%%%%%%%%%%%%%%%%%%%%%%%%%%%%%%%%%%%%%%%%%%%%%%%%%%%%%%%%%%%%%%%%%%%
%%%%%%%%%%%%%%%%%%%%%%%%%%%%%%%%%%%%%%%%%%%%%%%%%%%%%%%%%%%%%%%%%%%%%%%%%%%%%%%%%%%%%%%
\section{Implementation}
\label{sec:implementation}

In order to demonstrate BSMD as a distributed mobility information management system we implemented the BSMD nodes on \emph{Hyperledger} \footnote{\url{https://www.hyperledger.org}}\emph{Indy}, which is a \emph{public-closed} blockchain for decentralized digital identities and provides a framework for the exchange of information on secure \emph{peer-to-peer} connections. The lightweight libraries of \emph{Hyperledger Indy} makes it a good choice for the development of BSMD, where some of the transactions may have to be done using smartphones. %{It is worth noting that BSMD could be implemented in any other environment such as \emph{Hyperledger Iroha}}. 

\begin{figure}
	\centering
	\includegraphics[scale = .42]{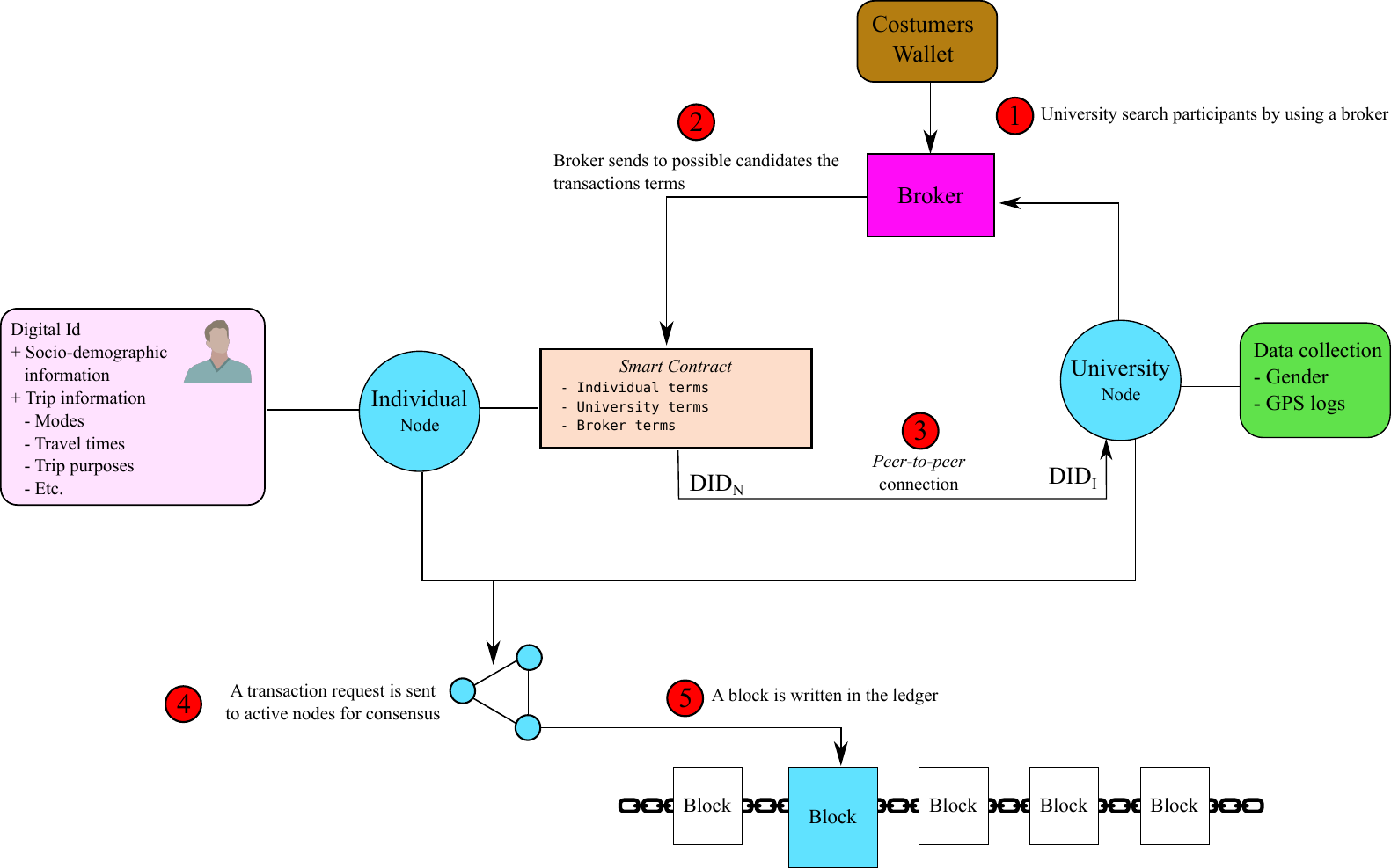}
	\caption{Transaction between two nodes in the BSMD}
	\label{fig:transactionIndy}
\end{figure}

In terms of the multi-layered model (\autoref{fig:layerBlock}), we implemented all the layers except the incentive layer, which can be added in future with minimum effort. Figure~\ref{fig:transactionIndy} shows with an example the details of operations implemented in BSMD. In this example we assume that there are three \emph{actives} nodes: University, Government agency, and a non-profit organization. They have the resources to participate in the consensus and to manage the ledger. The fourth and fifth nodes are \emph{passive} nodes associated to an `individual' and a \emph{broker}, who doesn't participate in these activities. {The University node will use BSMD for collecting mobility data during a four month period and it is willing to reward participants for its information.}

{In step $1$ the University contacts the \emph{broker} to get participants. University can contact the broker via mail, phone, and other usual business norms. For every transaction between the participants and University the \emph{broker} will charge some percentage ($10\%$ here) of the reward that will be enforced in the terms of the \emph{smart contract}. In step $2$ the \emph{broker} sends a transaction request to its clients. The transaction will have the terms of the University and a term to specify the transaction fees of the \emph{broker}}. 

{Algorithm~\ref{alg:smart_example} illustrates the terms of the \emph{broker}, University and an individual. In this case the \emph{smart contract} states that the information will be shared for a four month period. Hence a connection between the University and the Individual will be accessible during the predefined four month period only. Observe that University node is requesting mobility patterns and this information does not need to be transferred in real-time. The daily mobility patterns of the Individual can be sent during the night, when the bandwidth and device usage are minimal. Therefore, a connection will be opened late at night and when the information for that day is transferred the connection will be closed. This process will repeat during the four month period. Every time a transaction happens the \emph{smart contract} is triggered to enforce that the terms of all nodes are maintained and each party receives what they want.}

\begin{algorithm}
	
	\footnotesize
	\SetAlgoLined
	Individual = \{no service required ($s_o$), reward ($r_o$), differential privacy is use and gender ($p_o$), share mobility patterns from March to June ($t_o$), id key is required ($id_o$)\}\\
	University = \{no service provided ($s_r$), reward ($r_r$), low geographic accuracy and gender ($a_r$), share mobility patterns during for a four month period ($t_r$), no id proof key required $id_r$\}\\
	Broker = \{collect $10\%$ of $r_r$\}\\
	\SetKwProg{Fn}{Function}{}{end}
	\Fn{ShareInfo(Owner terms, Requester terms)}{
		\eIf{$s_o$ = $s_r$}{
			\eIf{$r_o \leq s_r$}{
				\eIf{$p_o \subset a_r$}{
					\eIf{$t_o \subset t_r$}{
						\eIf{University key pass verification}{
							Send $10\%$ of $r_r$ to broker\\
							start connection;
						}
						{
							refuse connection;
						}
					}{
						refuse connection;
					}
				}{
					refuse connection;
				}
			}{
				refuse connection;
			}
		}{
			refuse connection;
		}
	}
	\caption{\emph{Smart contract} Information transaction}
	\label{alg:smart_example}
\end{algorithm}

{Steps $3$, $4$ and $5$ will occur every day late at night. In step $3$ an encrypted \emph{peer-to-peer} channel of communication is opened by creating two Decentralized Identifiers ($DID$). The individual will create a $DID_I$ and send it to the University, while the University will create a $DID_N$ and send it to the individual. These $DID$s are the addresses that the nodes use for communication. Every time a connection is established a $DID$ is created, so a single node or individual can have multiple $DID$s assigned. In Step $4$ the nodes will submit a transaction request to the BSMD and an \emph{active} node will build a block with the characteristics of the transaction. This type of transaction is private and the information of the Individual node is never written in the ledger. Instead, the private information is encrypted and sent directly to the University node using the \emph{peer-to-peer} connection. Both nodes will keep the data in a private storage. BSMD can also accommodate public transactions where the the information is written in the ledger. For example, bike availability at stations is exposed by bike-share companies so users know in advance the nearest station with available bikes. The details on the block for these two cases are provided in \ref{sec:appedixblocks}.}

{More than one \emph{active} nodes can create their own block using a transaction. In Step $5$, by using a consensus protocol it will be decided, which \emph{active} node will write a block to the ledger.}

{The complete source code and instructions for using the BSMD implementation in \emph{Hyperledger} can be found at \url{https://github.com/billjee/bsmd}. Note that our framework is very flexible and generic. It can be implemented on various other blockchain SDKs beyond \emph{Hyperledger} with minimal effort.}
%We intent to show the benefits and potential of having a distributed ledger and transaction system where: (1) the privacy and security of individuals is maintained and (2) users can query all public transportation data in a unified platform. Later in Section~\ref{sec:discussion} we will discuss the potential benefits and challenges of BSMD framework. 

%%%%%%%%%%%%%%%%%%%%%%%%%%%%%%%%%%%%%%%%%%%%%%%%%%%%%%%%%%%%%%%%%%%%%%%%%%%%%%
%%%%%%%%%%%%%%%%%%%%%%%%%%%%%%%%%%%%%%%%%%%%%%%%%%%%%%%%%%%%%%%%%%%%%%%%%%%%%%
\subsection{Demonstrations and impact analysis}
{Distributed ledger technologies are an active area of research. However, to the best of our knowledge, it is rare to find studies outside the world of cryptocurrencies that have developed extensive analysis on the performance of their proposed frameworks. Table~\ref{tab:blockchainSimulations} lists the major studies on this matter. Majority of them have been small-scale simulation studies, focusing on the technical aspects only. Moreover, to the best of our knowledge, no such general framework currently exists for a blockchain based mobility information management framework.}

\tabulinesep=1.5mm
\begin{table}[!h]
	\small
	\caption{Blockchain simulations}
	\label{tab:blockchainSimulations}
	\begin{tabu} to \linewidth  {| X[l,m] | X[l,m] | X[l,m] | X[,l,m] | X[l,m] |}
		%%\begin{tabular}{m{6em} m{5em} m{5em} m{5em} m{5em} m{5em} }
		\hline
		\textbf{Framework} & 
		\textbf{Description} & 
		\textbf{Setup} & 
		\textbf{Size of the blockchain} &
		\textbf{Test}\\
		\hline
		Eikeden~\cite{Cheng2018} & 
		Privacy-preserving \emph{smart contracts} & 
		Master: One local i7, 8GB RAM machine. \newline Slave: 4 EC2 \emph{t2.medium} & 
		4 nodes & 
		Latency and throughput \\
		\hline
		Hyperledger Fabric~\cite{Sousa2018} & 
		Business blockchain framework & 
		\begin{itemize}[leftmargin=*]
			\item[a.] Local Xeon E5520 32GB RAM server. \item[b.] 4 EC2 \emph{m4.4xlarge}
		\end{itemize}
		& 
		\begin{itemize}[leftmargin=*]
			\item[a.] 10 nodes in local simulations. \item[b.] 4 nodes in distributed simulations
		\end{itemize}
		& 
		Latency, throughput, RAM and HHD usage, size of transactions, impact of CPU in nodes \\
		\hline
		BaDS~\cite{Zhang2018b} & 
		Blockchain framework for IoT & 
		1 local Core i7, 8GB RAM & 
		Not reported & 
		Computational cost of transactions \\
		\hline
		BeeKeeper~\cite{Zhou2018} & 
		Blockchain-enabled IoT & 
		4 1GHz CPUs, 2GB RAM servers & 
		Not reported &
		Latency and throughput \\
		\hline
		Smart-Grid~\cite{Mengelkamp2018} & 
		Blockchain for trading local energy generation & 
		Not reported & 
		Not reported &
		Market prices \\
		\hline
	\end{tabu}
\end{table}

The studies presented in Table~\ref{tab:blockchainSimulations} either tested their blockchain on a local computer or on a combination with Amazon Elastic Compute Cloud (EC2). For the evaluation of the performance, latency and average throughput are the common denominators in these simulations. Hence, in order to keep simulation features at least at the same level as that of state-of-the-art, we developed two demonstrations and analyzed their performance using latency and average throughput.

The first demonstration simulated the management of real-time mobility data (e.g. GPS stream) on BSMD, generated over 24hr by mobile devices e.g. smartphones. Latency and throughput were analyzed for 106-206 nodes implemented on BSMD. The simulation ran on an Ubuntu 16.04 machine with an i7-8cores (3.5GHz) and 16GB of RAM. Note that the blockchain size here is larger than most of the studies presented in Table~\ref{tab:blockchainSimulations} where a simulation is performed on a local machine to test the blockchain in a controlled environment without the performance impacts of external factors e.g. available bandwidth on a communication network.

The setup of the second demonstration is more comprehensive both physically as well in terms of the size. Here the real-time mobility data management is simulated on a physical network and with a blockchain composed of 370 BSMD nodes that were geographically-separated and were using different communication networks:
\begin{enumerate}
	\item[a.] 10 \emph{t2.medium} (Amazon cloud EC2 virtual machines with 2cores at 3.1GHz and 4GB of RAM), running 1 \emph{active} node each
	\item[b.] 20 Raspberry Pi (RPi) model 3B (4cores at 1.2GHz and 1GB RAM) using WiFi and running 10 \emph{passive} nodes each
	\item[c.] 1 local computer (i7-8cores at 3.5GHz and 16GB of RAM) using Ethernet for communication, running 20 \emph{passive} nodes in each core
\end{enumerate}

The EC2 computers ran the \emph{active} nodes i.e. those who participate in the consensus mechanisms and write block in the ledger. On RPis and local computer, individual nodes were implemented that were sharing their mobility information on the blockchain. This demonstration is a very close approximation of the reality where users share their mobility information in the blockchain using their hand-held devices because (a) RPi computing capabilities are a close approximation of a mobile device and (b) simulation running on a real infrastructure where WiFi signals, speed of the internet, and type of device may also affect the performance. 

Besides the standard operations of BSMD (see Figure \ref{fig:transactionIndy}), following operations specific to mobility data sharing scenario are also implemented: (a) \emph{active} nodes validating other nodes so that individuals (e.g. mobile devices) can trust them when sharing information (b) creation of \emph{smart contracts} specific to the mobility data sharing on BSMD (c) \emph{peer-to-peer} encrypted  communication for sharing mobility information. Next, we present the implementation details of the experiments and results of both demonstrations.

\subsubsection{BSMD simulation on local machine}
\label{sec:localSimulation}
To run BSMD on a local machine we installed 6 \emph{active} nodes on 3 cores and used the remaining 5 cores to simulate different sizes of mobile device populations (\emph{passive} nodes). Distributing the BSMD nodes on cores simulate a real case where the nodes interact in the blockchain using different devices. In total we simulated populations of $100$, $125$, $150$, $175$ and $200$ mobile devices. Our local computer was not able to simulate populations beyond $200$ and started crashing, so no further simulations were performed.

%\textcolor{blue}{The sharing of GPS information involved the following steps:}
%\begin{enumerate}
%    \item[a.] \textcolor{blue}{The sender created DID for the communication; 
%    \item[b.] The sender encrypts a connection; request with their DID and a verification key to identify the receiver;
%    \item[c.] The sender decrypt a connections response the sender DID and their verification key;
%    \item[d.] If the connection is successful i.e. the sender can decrypt the connection, a successful connection is written in the ledger;
%    \item[e.] The nodes share data via a \emph{peer-to-peer} connection.}
%\end{enumerate}

In reality, population share real-time location data throughout the day, so the simulation ran for a 24hr period. Assuming that an individual sends location points every 5sec, so in 24hr the maximum number of GPS points their devise (e.g. smartphone) could share is $17,280$. On the other extreme, if the individual stay at home all day the minimum number of points shared is $0$. In general most of the individuals will share their home-work trips and errands/entertainment.  {We assumed a more extreme scenario where individual users are sharing their location for the most part of their day (even if they are stationary). Hence at an average, an individual will share approximately $9,300$ GPS points throughout the day.}
% previously we had write that 
% "that the daily total travel time of an average individual is $2.5$ hours, i.e., at . an average individual will share approximately $9,300$ GPS points throughout the day"
% which is totally wrong because 2.5hours of points is 1800 

\begin{figure}[!h]
	\centering
	\includegraphics[width=0.75\textwidth]{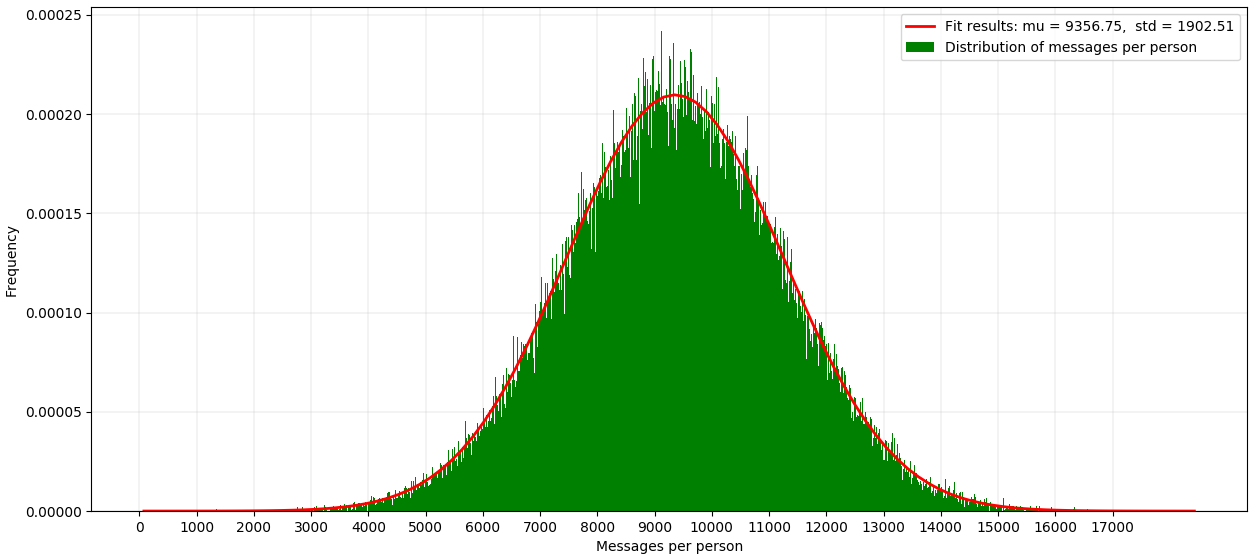}
	\caption{Distribution of real-time locations in a population of $100$ individuals}
	\label{fig:msg_distribution}
\end{figure}

Based on that, for a population of $n$ individuals we assumed that the number of real-time locations generated by individuals is normally distributed with mean, $\mu = 9356$, and standard deviation, $\sigma = 1902$. Figure~\ref{fig:msg_distribution} shows an example of the distribution of real-time locations generated by $100$ individuals.

\begin{figure}
	\centering
	\includegraphics[width=\textwidth]{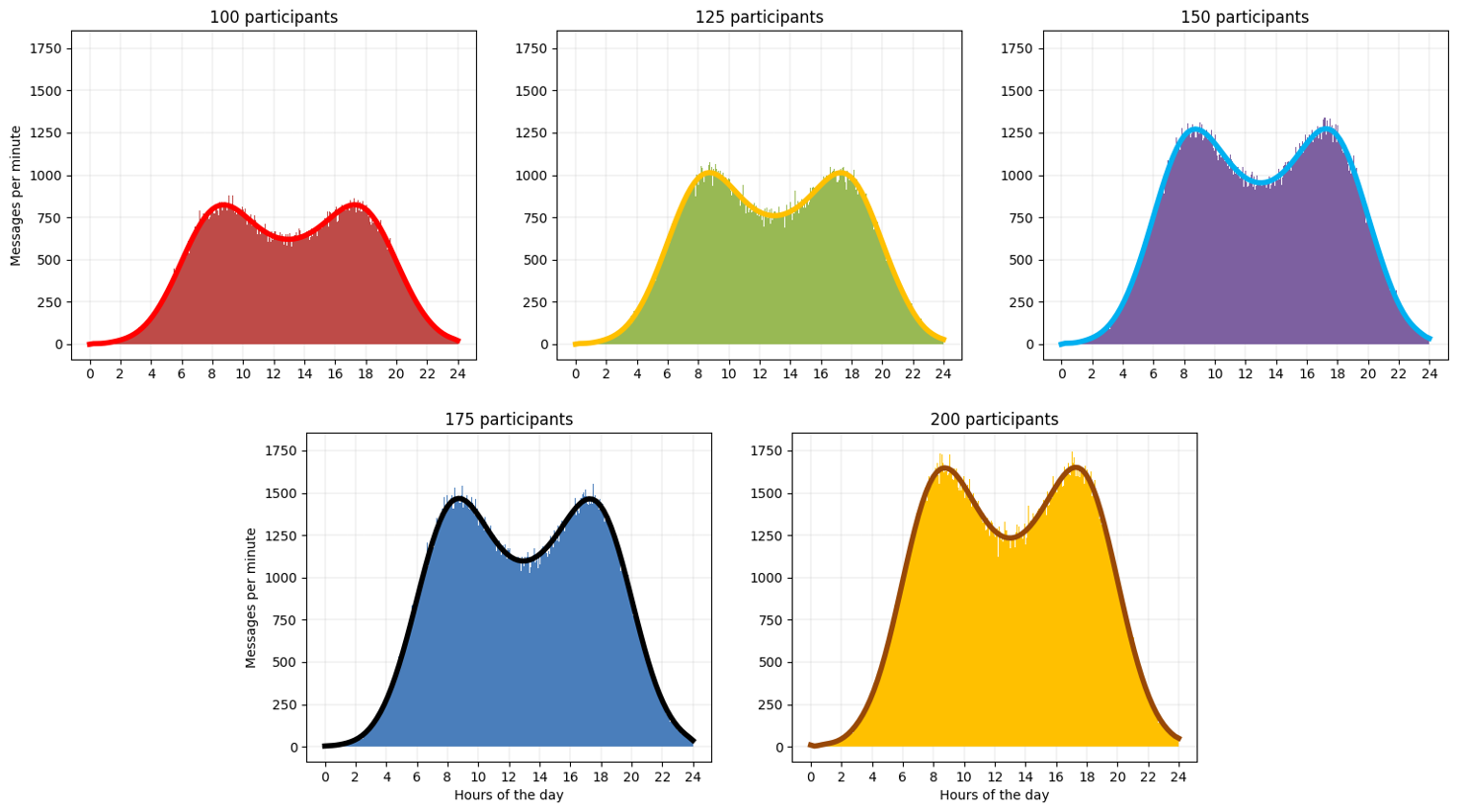}
	\caption{Distribution of messages during the day for populations of $100$, $125$, $150$, $175$ and $200$ individuals}
	\label{fig:pop_distribution}
\end{figure}

Assuming that most trips are around 08:00 (morning peak) when people go to work and around 18:00 (afternoon) when people return to home, the distribution of the messages a population sends during the day is simulated with a mixture of three normal distributions, such that $\mu_1 = 8$ and $\sigma_1 = 2.3$ represent home to work trips, $\mu_2 = 13$ and $\sigma_2 = 3.5$ represent lunch trips and $\mu_3 = 18$ and $\sigma_3 = 2.3$ represent work to home trips. Figure~\ref{fig:pop_distribution} shows the distributions of messages sent by populations of $100$, $125$, $150$, $175$ and $200$ individuals. Table~\ref{tab:blockchainCharac} we present the main characteristics of each experiment.

%\tabulinesep=1.5mm
\begin{table}
	
	\footnotesize
	\caption{Main characteristics of each simulation}
	\label{tab:blockchainCharac}
	\begin{tabu} to \linewidth  {| X[c,m] | X[c,m] | X[c,m] |  X[c,m] |}
		%%\begin{tabular}{m{6em} m{5em} m{5em} m{5em} m{5em} m{5em} }
		\hline
		\textbf{Population Size} 
		& \textbf{Total messages} 
		& \textbf{Maximum locations shared / min.} 
		& \textbf{Average locations shared / min. (st. dev)}\\
		\hline
		$100$ & $944,678$ & $1,160$  & $878$ ($\sigma = 233$) \\
		\hline
		$125$ & $1,174,204$ & $1,418$ & $1,071$ ($\sigma = 281$) \\
		\hline
		$150$ & $1,360,010$ & $1,746$ & $1,318$ ($\sigma = 356$) \\
		\hline
		$175$ & $1,662,254$ & $1,963$ & $1,517$ ($\sigma = 405$) \\
		\hline
		$200$ & $1,857,092$ & $2,296$ & $1,761$ ($\sigma = 484$) \\
		\hline
	\end{tabu}
\end{table}

% 100 population
% Total msg	67050406
% max msg	1160
% avg msg	878.750308
% std	233.0733126

% 125 population
% Total msg	100715821
% max msg	1418
% avg msg	1071.125845
% std	281.6361953

% 150 population
% Total msg	143923069
% max msg	1746
% avg msg	1318.061313
% std	356.5305323

% 175 population
% Total msg	195153237
% max msg	1963
% avg msg	1517.6629
% std	405.0956754

% 200 population
% Total msg	241537459
% max msg	2296
% avg msg	1761.607
% std	484.8313377

The results of the simulations are shown in Table~\ref{tab:blockchainSimuPop}. In general, the latency remains under $0.6$ secs even when the system is high congested with the messages. Thus the response times of the BSMD are promising to keep up with the demands of the the modern world. In actual implementations, this response can be improved further by deploying more active nodes with higher computational power.

The throughput shown in Table~\ref{tab:blockchainSimuPop} is the ratio of message served and messages send i.e. an average throughput of $0.9$ mean that $90\%$ of the messages were processed. It is expected that the average throughput decays as the population grows because the number of cores in the local computer hosting the blockchain and the population remains constant. It is worth noting that in the simulation for $200$ individuals the core and RAM usage was at $100\%$ through the peak hours (from minute $480$ to $1080$) of the experiment which may explain why some messages were lost. In general the throughput remains 90\% or more for all the experiments on a limited hardware, which shows a high reliability of the system.

\tabulinesep=1.5mm
\begin{table}[!h]
	\small
	\caption{Blockchain performance with different populations}
	\label{tab:blockchainSimuPop}
	\begin{tabu} to \linewidth  {| X[0.4,c,m] | X[c,m] | X[c,m] |}
		%%\begin{tabular}{m{6em} m{5em} m{5em} m{5em} m{5em} m{5em} }
		\hline
		\textbf{Population} & \textbf{Average Latency in sec.} & 
		\textbf{Average throughput}\\
		\hline
		100 & $0.014$ $(\sigma = 0.006)$ & $1$ $(\sigma = 0)$ \\
		\hline
		125 & $0.14$ $(\sigma = 0.41)$ & $0.976$ $(\sigma = 0.134)$\\
		\hline
		150 & $0.33$ $(\sigma = 0.64)$ & $0.943$ $(\sigma = 0.135)$\\
		\hline
		175 & $0.52$ $(\sigma = 0.62)$ & $0.937$ $(\sigma = 0.152)$\\
		\hline
		200 & $0.60$ $(\sigma = 0.76)$ & $0.9$ $(\sigma = 0.145)$\\
		\hline
	\end{tabu}
\end{table}

\begin{figure}
	\centering
	\includegraphics[width=0.63\textwidth]{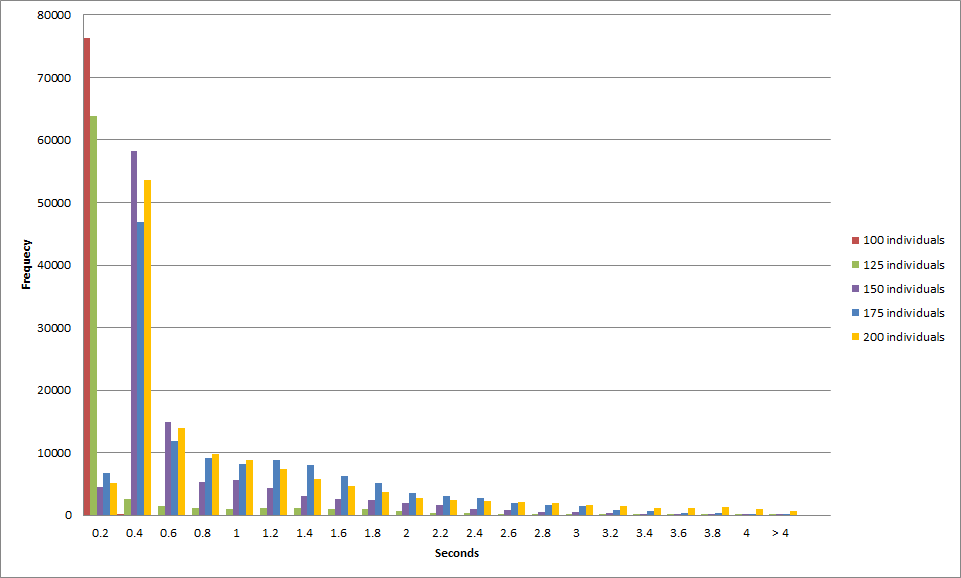}
	\caption{Latency frequency in seconds for populations of $100$, $125$, $150$, $175$ and $200$ individuals}
	\label{fig:latency_distribution}
\end{figure}

Furthermore, we split the analysis of the latency into two parts. First the frequency of latency of a single message is analyzed and then in the second part we show how the average latency changes through the day. Figure~\ref{fig:latency_distribution} shows the frequency of the latency for the populations. We observe that the latency of all messages remains under $0.2$sec. for a population of $100$ individuals. While for the populations of $125$, $150$, $175$ and $200$ the percentages of messages with latency under 1sec are $91\%$, $85\%$, $70\%$ and $70\%$ respectively. For populations over $100$ the frequency of the latency decay abruptly from $0$ to $0.6$sec, and after $0.6$sec the frequency seems to decay slowly at a constant rate.
% 200 wallets 30% over one second
% 175 wallets 30% over one second
% 150 wallets 15% over one second
% 125 wallets 9% over one second

\begin{figure}[!h]
	\centering
	\includegraphics[width=0.63\textwidth]{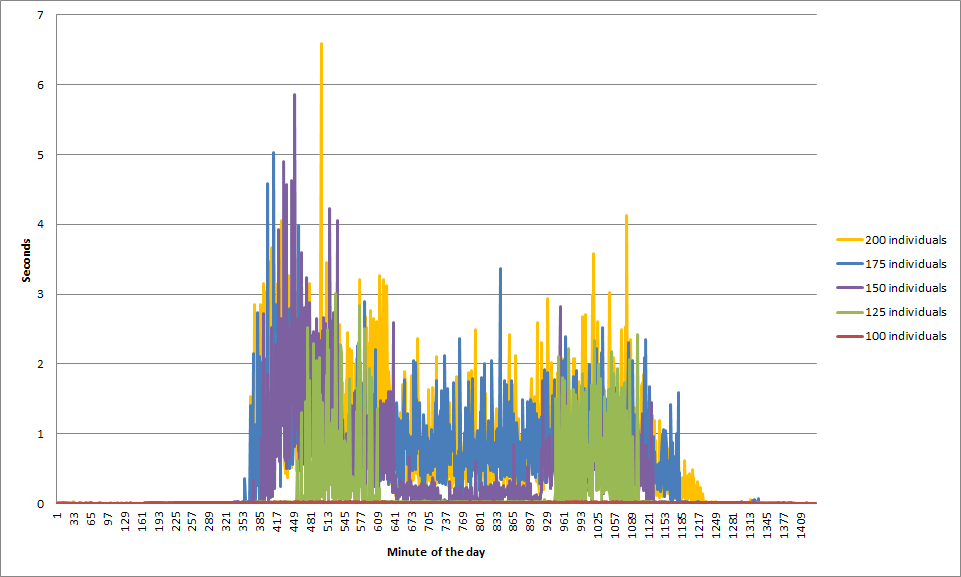}
	\caption{Average latency per minute of the day of the different populations}
	\label{fig:avg_distribution}
\end{figure}

\begin{figure}[!h]
	\centering
	\includegraphics[width=0.63\textwidth]{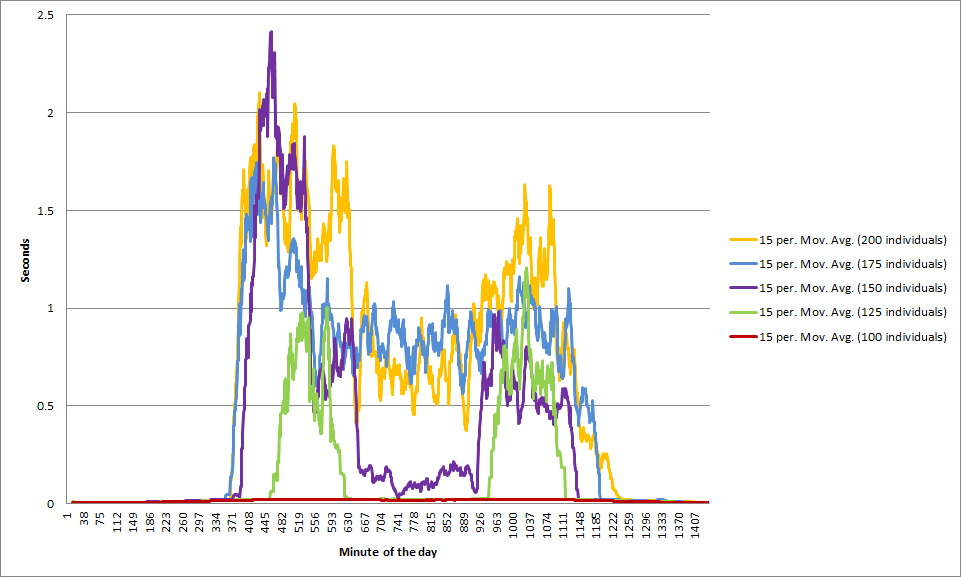}
	\caption{15 minute moving average of the average latency per minute of the day}
	\label{fig:moving_avg}
\end{figure}

Figure~\ref{fig:avg_distribution} shows the average latency per minute over a day for all the populations. Since the maximum latency of the $100$ population is less than $0.028$sec, the average latency per minute is presented as a flat line. As expected, the latency increases around the peak hours (08:00 and 18:00) and in between the peak hours the latency decays. Although all populations present spikes for few minutes of the day, if we eliminate the noise by considering the moving average of the 15 previous minutes (see Figure~\ref{fig:moving_avg}) then the latency for all the experiments is under $2.5$sec and for most part of the day the latency is almost under $1.5$sec.

% Persons	AVG	DESV
% 100	1.000	0%
% 125	0.976	0.134173541
% 150	0.943	0.135900964
% 175	0.937	0.152398507
% 200	0.900	0.145797092

% Average	        0.014435166	0.141147064	0.335548182	0.522696206	0.604364226
% Standar deviation	0.006530729	0.419581434	0.640074635	0.623361638	0.767800234
%                 	100         	125     	150     	175 	200
\subsubsection{BSMD simulation on a physical network}
\label{sec:DisSimulation}
Here we simulated a blockchain of 370 nodes on a physical network. We used 10 BSMD \emph{active} nodes installed on 10 \emph{t3.medium} EC2 virtual machines. We also used 20 RPis and one local machine to simulate a population of (10 nodes x 20 RPis) + (20 nodes x 8 cores) = 360 \emph{passive} nodes sharing their real-time location data. Given the low hardware specifications of the RPis and the fact that it is necessary to encrypt and decrypt messages in the blockchain, we could only test a population of 10 individuals per RPi. In this simulation we assumed that each individual is sharing real-time locations during the morning peak-hour, from 08:00 to 09:00.

Just like Section~\ref{sec:localSimulation}, here too we assumed that the individual device is sending location points every 5sec. We used a normal distribution with $\mu=9356$ and $\sigma = 1902$ for generating the real-time locations of each individual. And we use a mixture of the distributions $\mu_1 = 8$ and $\sigma_1 = 2.3$, $\mu_2 = 13$ and $\sigma_2 = 3.5$ and $\mu_3 = 18$ and $\sigma_3 = 2.3$ for generating all the GPS points the population share in the blockchain. Hence, during the morning peak hour the \emph{passive} nodes on RPi and the local computer shared $11,443$ locations with the \emph{active} nodes on \emph{t3.medium} virtual machines.

Table~\ref{tab:blockchainRPICUP} shows the results of the simulation. Given that each individual sends a message every 5sec, it is desirable that the latency is below this time so as to maintain the real-time flow of information. In Table~\ref{tab:blockchainRPICUP} the average latency of the local machine simulation is $2.9$sec, however for the RPis it is $6.59$sec. This difference is due to the low computation power of RPis and because each of them is simulating 10 nodes simultaneously. Although our tests showed that an RPi can handle the decryption and encryption of the connections request/response for a single node with a latency of less than 1sec, it consumes all of the RPi resources to perform this task for 10 nodes simultaneously. The low latency times are thus more related to the computational power of the RPi than the time needed to get a response from the blockchain. In real-life, mobile devices like smartphones would only run 1 node and would also be more powerful than an RPi we used e.g. iPhone 7, which is an older generation smartphone (released in 2016) has 4 cores at 2.34 GHz and 2GB of RAM. The average throughput in this experiment for both RPis and the local machine was $99\%$. This demonstrates the high reliability of the implemented BSMD. Despite a relatively higher latency exhibited by the overloaded RPis, there is a very low rate of message loss and most of the messages are successfully being sent and received by \emph{active} and \emph{passive} nodes in the blockchain.

\tabulinesep=1.5mm
\begin{table}
	\small
	\caption{Blockchain performance}
	\label{tab:blockchainRPICUP}
	\begin{tabu} to \linewidth  {| X[c,m] | X[c,m] | X[c,m] |}
		%%\begin{tabular}{m{6em} m{5em} m{5em} m{5em} m{5em} m{5em} }
		\hline
		\textbf{Population} & \textbf{Average Latency in sec.} & 
		\textbf{Average throughput}\\
		\hline
		20 RPi, 10 users per RPi             & $6.59$ $(\sigma = 2.32)$ & $0.998$ $(\sigma = 0.008)$ \\
		\hline
		Local machine, 20 users per core       & $2.98$ $(\sigma = 0.82)$   & $0.991$ $(\sigma = 0.021)$\\
		\hline
		RPi + Local Machine & $5.56$ $(\sigma = 1.73)$   & $0.996$ $(\sigma = 0.009)$\\
		\hline
	\end{tabu}
\end{table}

%   	    latency	    troughput	latency RPi	trougput RPi	lat CPU 	trougpu CPU
% avg	    5.560950375	0.996138813	6.592762609	0.998184229	2.981419789	0.991025274
% st dev	1.733725027	0.009251125	2.325039305	0.008911836	0.82011752	0.021488136

\subsection{Cybersecurity and privacy impacts}
BSMD addresses some key cybersecurity and privacy issues individual user faces these days, while sharing mobility data information. When sharing, it is not easy for the individual to tell if the receiver is actually the one who it claims it is. This situation can result is \emph{spoofing} where fake companies/entities can trick users into sharing information for malicious activities. By using the BSMD implementation, it is possible to verify the identities in blockchain. The identity key system can be managed by all the trusted nodes (government or non-profit organizations). 

In case of \emph{message interception} attack, the individual's communication with a legitimate entity can be intercepted by a malicious entity. The peer-to-peer encrypted communication in BSMD makes it very difficult for malicious entities to launch such attack.

The individual users would like to have the ability to stop sharing their information at any point or bar a node for using their information further. \emph{Smart contract} implemented in BSMD provides the mechanism to give this control to the individuals.

Next we present the details on how BSMD safeguards against \emph{spoofing} attacks. However, the detailed implementation of all the three cases i.e. \emph{spoofing}, \emph{message interception}, and \emph{privacy control} can be found at: \url{https://github.com/billjee/bsmd/tree/master/examples}. 

\subsubsection{Spoofing attack}
\begin{figure}
	\centering
	\begin{subfigure}{.5\textwidth}
		\centering
		\includegraphics[scale = 1.2]{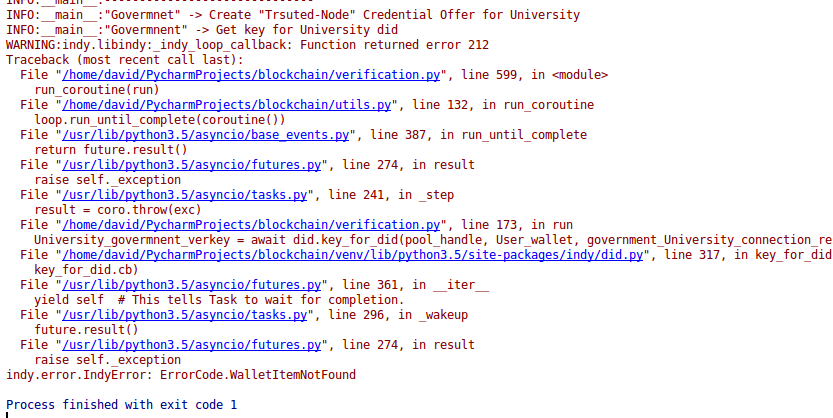}
		\caption{Without application proof}
		\label{fig:spoofing1}
	\end{subfigure}\hfill
	\begin{subfigure}{.5\textwidth}
		\centering
		\includegraphics[scale = 1.2]{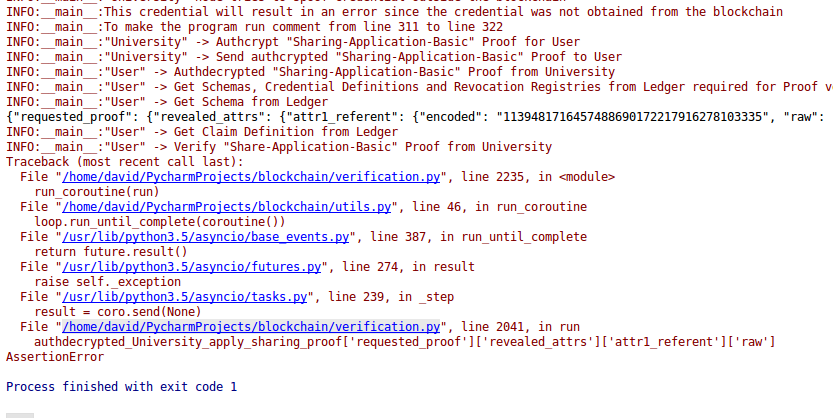}
		\caption{With application proof}
		\label{fig:spoofing2}
	\end{subfigure}
	\caption{Node attempting to access Individual's identity}
	\label{fig:spoofing}
\end{figure}

Here we use the same mobility case study as described in Section \ref{sec:localSimulation}. Assume that the six \emph{actives} nodes are: Federal Government, Provincial Government, City Government, University, a RideHailing Company, and a Non-For-Profit Transportation Organization. They have the resources to participate in the consensus and to manage the ledger. We also chose one random \emph{passive} node `Individual' from the population, who does not participate in these activities but only share their GPS data. If the University wants to get mobility data from the Individual first the Individual needs to verify that the Identity of University is approved by any of the Government node. In order to get a identity key from the Government the following steps are required:

\begin{enumerate}
	\item Government creates a credential schema and sends this definition to the ledger. Schema is a blank credential that the Government will use for creating identity keys.
	\item Government issues a credential with the identity key of the University.
	\begin{enumerate}
		\item University stores the credential in its wallet.
		\item A registry of the credential is sent to ledger. This registry will be used by others to verify the identity of the University.
	\end{enumerate}
	\item University makes a connection request to the Individual for sharing information
	\item Individual sends the University a sharing application.
	\item University fills this application with the identity key issued by the Government. Using the application filled, the University obtains from the ledger an application proof that will be used by the Individual node for verifying the identity and establishing the connection. If the University tries to spoof Individual's identity the registry from the ledger will not match with the application and hence the University will not get the application proof making the connection with the user impossible. 
	Figure~\ref{fig:spoofing1} shows the output log when the University tries to spoof Individual's identity before obtaining application proof.
	
	\item Individual reads the application and compares it with the registry in the ledger. If application and the registry match, Individual can validate the identity of the University node. If the application and the registry does not match the Individual rejects the connection as the Identity of the node may be spoofed. Figure~\ref{fig:spoofing2} shows the output log when the University node tries to access Individual's identity after obtaining application proof. 
\end{enumerate}

In the presented example, there are two steps of verification of the identity. In first step, the University tries to get an application proof for connection, if the application proof does not match with the registries of the ledger the application proof for connection is not proceed. In second step, if the University tries to access the Individual's identity after obtaining the application proof for connection. In this step the Individual will also compare the received information with the registries from the ledger to verify the identity of the University. 

The two step verification process makes it harder for nodes to spoof Individual's identity because it is required to compare node's information with the registries of the ledger. One comparison is made by the node itself and the other is made by the node with whom it wants to connect to. The only case where a node can spoof someone's identity would be to tamper the ledger, which is practically impossible due to the design of BSMD. Please also note that all these processes would be implemented in software and would not bring any extra burden on the actual users themselves. In most cases, these features will be turned on by default on the BSMD enabled devices they are using.

%%%%%%%%%%%%%%%%%%%%%%%%%%%%%%%%%%%%%%%%%%%%%%%%%%%%%%%%%%%%%%%%%%%%%%%%%%%%%%%%%%%%%%%
\section{Discussion}
\label{sec:discussion}

Nowadays, the flow of the mobility information can be summarized as following steps: 
\begin{enumerate}
	\item Mobility information is generated by an individual. As they move, details such as mode, travel time, departure time, etc, can be recorded by their smartphones~\cite{Yazdizadeh2018} or can be described by themselves.
	\item The information is solicited actively or passively. Third parties can ask for mobility information using surveys or by getting access to sensors data of an individual's smartphones. 
	\item Information is collected. The collection methods can be via phone, in person, online, or actively collected using smartphones.
	\item The collector processes and analyzes the information. The collected data is processed and stored in central databases from where data analysis can be performed.  
\end{enumerate}

The six layers of BSMD are developed in such a way that they include the steps 1 to 3 in a single environment, making the transaction of mobility information secure, private and in some cases, the collection of mobility information is easier than the current practices. The mechanisms for processing and aggregation of information are not part of BSMD because each collector will choose different tools for these tasks. However, the collector can use the blockchain for sharing or selling the collected information.

The privacy and identification layers are for the generators of information (step 1). The privacy layer enforces the principles on which a BSMD is built upon and a model which maintains location privacy of the users. The identification layer has the tools for collecting mobility data. The identification can be build in a standardized way (e.g. json format) so that the collectors of information can easily process and analyze the data. {In this layer nodes can store identity keys, which can be used by other nodes to verify them.}

The contract, incentive, communication and privacy layers are for soliciting information (step 2). These layers have the mechanisms for selecting which information individuals are willing to share and a secure way for data transfers. {\emph{Smart contracts} are used to self-enforce fair trade and to manage disputes, such as breaking the terms of transactions or price negotiation}. Using incentives, the nodes can convince individuals to share their information.

The collection of information (step 3) uses all the BSMD layers. A collector node will use the identification layer to get the information from other nodes. This information can only be accessed using the contract and privacy layers. The collector can use the incentive layer to encourage others to disclose their information. All information is transferred using the communication layer. Once the transaction is completed the consensus layer will check if the transaction is valid and a block is written in the ledger.

\subsection{Opportunities}
BSMD aims to drastically improve various aspects in the current way of sharing and collecting mobility data. First, it will give back the generator of the information full control over their information and privacy. Nowadays, it is not easy or in most cases, entirely not possible to access certain services unless users consent to all the conditions imposed by the collectors. Most of the time, the collectors have full control over users' information. \emph{Smart contracts} and public ledger will help users protect, control and track their information. Every transaction in BSMD needs to be validated by other nodes using consensus mechanisms. Thus, unsolicited request of information can be prevented.

{When nodes sell information to a third party they do not lose total control over their information, because it is legally protected by terms of the \emph{smart contract}. If the requester node is caught using the information outside the terms of the \emph{smart contract}, then the auditors can take advantage of the public ledger to see which nodes were scammed and take legal actions against the one responsible. Having a public and distributed ledger may speed up the legal process since all transactions are accessible. Nowadays, when companies are caught misusing information, it is hard to prove whose information was misused, since all records are on private servers. Also, it is difficult for authorities to audit private servers to measure the true extent of the damage.}

Creating a unified platform where mobility information can be shared in a secure and transparent way will improve the collection of information and hence the analysis of the data. Nowadays, when companies/governments share mobility information they create APIs and develop security mechanisms for others to access their information. Despite the existence of standards for developing such APIs, companies/governments need to spend resources on the development. The blockchain will provide a single and secure network where companies/governments just need to ``plug-in'' and start sharing and accessing mobility data.

\subsection{Challenges}
Blockchain technology is in its early stages of development and there are a number of challenges that need to be addressed for making BSMD readily available. The monetization of users mobility information is a big business for companies, so further research on monetization schemes needs to be developed in order to convince companies to adopt BSMD. {In Section~\ref{sec:incentive}, we developed a game theory based mechanism to analyze the case where companies can buy information and still make a profit. However, it is not clear what is the fair price for the mobility data, where companies can still profit from individual's mobility information and at the same time protect their privacy. An open-market based on the demand can be a good solution. The design and development of such market could be an interesting topic for a future work.} 

Although the percentage of users with smartphones or access to computers grows every day, the ubiquitousness of this technology is still few years away. For instance, in 2015 smartphone ownership in the world was $46\%$, while in richer economies this rate was up to $88\%$~\cite{Pew2016}. It is possible that some analysis of the information collected in BSMD will marginalize certain sectors of the population, making their mobility issues invisible to society. When all sectors of the population have a representative sample  of smartphone ownership more robust analysis of the mobility information collected in BSMD will be performed. For the moment, transportation studies using the data collected in BSMD will be more useful for populations with high smartphone ownership.

{There is a trade-off between \emph{public} and \emph{close} blockchains. In the former, anyone can participate in consensus and host the ledger. However, the consensus requires large use of energy as well as the capacity to handle several transactions per second. BSMD is of the latter type. Here, the participants need approval from \emph{active} nodes to participate in the consensus and hosting the ledger. This type of blockchain can handle considerably more transactions than its counterpart and requires less energy for consensus. In order to keep the BSMD as democratic as possible, the power that the \emph{active} nodes have over BSMD must be legally regulated. Rules need to be set and enforced so when users wants to become an \emph{active} node, they will have the means and will to do that. Defining the set of rules and the mechanisms imposed on the \emph{active} nodes is out of the scope of this paper.}

{Theoretically, in \emph{public} blockchains like \textit{Bitcoin} anyone can mine blocks, participate in consensus, and host the ledger. In reality though, only the people/companies who have large computing resources stay active. The cheapest mining computer that can make profits of a couple hundred dollars a year can cost as much as \$2,000. Since \emph{closed} blockchain requires considerably less computing power, money may not be a problem and people will able to easily buy computers that can participate in the consensus mechanisms.}

{BSMD is able to automatically resolve disputes and enforce fair trade with \emph{smart contracts}. By using $DID$s and asymmetric encryption in \emph{peer-to-peer} connections, communication jamming can be prevented. The identity keys will reduce the number of fake users and node spoofing attacks. With identity key, a node can verify sector, address and company information of governments/agencies/organizations, as well as gender and age of real persons. At the moment BSMD cannot be used to prevent malicious node from selling spoof location data. Detection of spoof GPS location data is still an open problem and may require invading the privacy of users~\cite{Zhao2018a}. For instance,~\cite{Zhao2017} proposed digging into the social media of users and to detect the probability that the user is in a certain location. The solution by~\cite{Tian2017} needs the users to send photos of landmarks to prove location. In the authors opinion and to keep consistency with the nature of BSMD, privacy should be the first priority and any solution that solves the fake user problem by affecting the privacy of individuals should not be taken into account.}

Scalability is another aspect that is worth considering. For BSMD to be successful it has to be capable to manage tens of thousands of transactions per second. One limitation of the blockchains that use proof-of-work for consensus is that at most it can handle 100 transaction per second and latency of 1 hour. So despite its strong security features, this type of consensus mechanism cannot be used in BSDM. SMART, a variant of Byzantine Fault-Tolerant (BFT) can handle up to 2300 transaction per second and has a latency of half a second~\cite{Sousa2017}. \emph{Hyperledger Fabric}~\cite{Androulaki2018} can achieve 3500 transactions per second using special configurations. Nevertheless the security of BFT-SMART and \emph{Fabric} are guaranteed as long as only trusted nodes participate in consensus. Given that BSMD is \emph{public closed}, BFT-SMART may be a solution for fast and secure transactions in big networks.

\subsection{Privacy principles}
BSMD is build upon the principles of:
\begin{enumerate}
	\item Data Ownership
	\item Fine-grained Access Control
	\item Data Transparency
	\item Auditability
\end{enumerate}

The main objective of BSMD is to give back to users the power to control and profit from their transportation data.

\subsubsection{Data ownership}
\label{sec:DataOwnership}
Most people do not completely read privacy policies or terms and conditions forms, they just accept these forms to enter web sites~\cite{Obar2016}. Selling information in BSDM is a good incentive for users to guard their privacy and read terms and conditions as this may imply that users will not give their information before checking if the profit they make is a fair exchange. The more information they sell the bigger the reward will be, but at the same time their privacy will be reduced.

In BSMD, people have the right to be forgotten. All transactions of information are encrypted and the only way to decrypt the information is if a connection is open between the nodes. So, when a node shuts down a connection the information that was shared will not be accessible by the receiving node because the description key will no longer be available.

\subsubsection{Data transparency and auditability}
\label{sec:DataTransparency}
There have been cases where companies share personal data to untrusted parties~\cite{Solon2018} without the owner's consent.~\cite{Bonneau2009} showed that the protection of personal data is not at the same level as that of user's expectations. The ledger in BSMD is open to the public so anyone can track their personal transactions by using their DID. {If a misusing node shares information and breaks the terms of the \emph{smart contract} authorities and owners can query the ledger to see which information they shared. Thus, the affected parts in the unwanted sharing of information can be easily tracked and legal actions can be taken, since all nodes have the proof that their information was misused}.   

\subsubsection{Fine-grained access control}
\label{sec:FineGrained}
Currently, on-line services enforce an all-or-nothing model i.e. when users subscribe to services, it is a common practice to make them accept all the terms and conditions for sharing all the data a service wants. If a user does not want to agree on sharing parts of the data, their only option is not to use such services. By using \emph{smart contract} the BSDM can get rid of such extreme cases and there can be levels of services/rewards for the level of data one shares.

Nowadays, companies store mobility data and individuals cannot revoke access to some or all information unless they opt-out and stop receiving the service. In BSMD \emph{smart contract} may have a clause to revise the information a user wants to share after a certain time, for example, after a year a user may update the terms in the contract for sharing information. In BSMD instead of companies setting the terms of use, the user will set those terms and companies will have to work according to each individual's terms of use. 

\subsection{Adversaries}
\begin{figure}
	\centering
	\includegraphics[scale = .25]{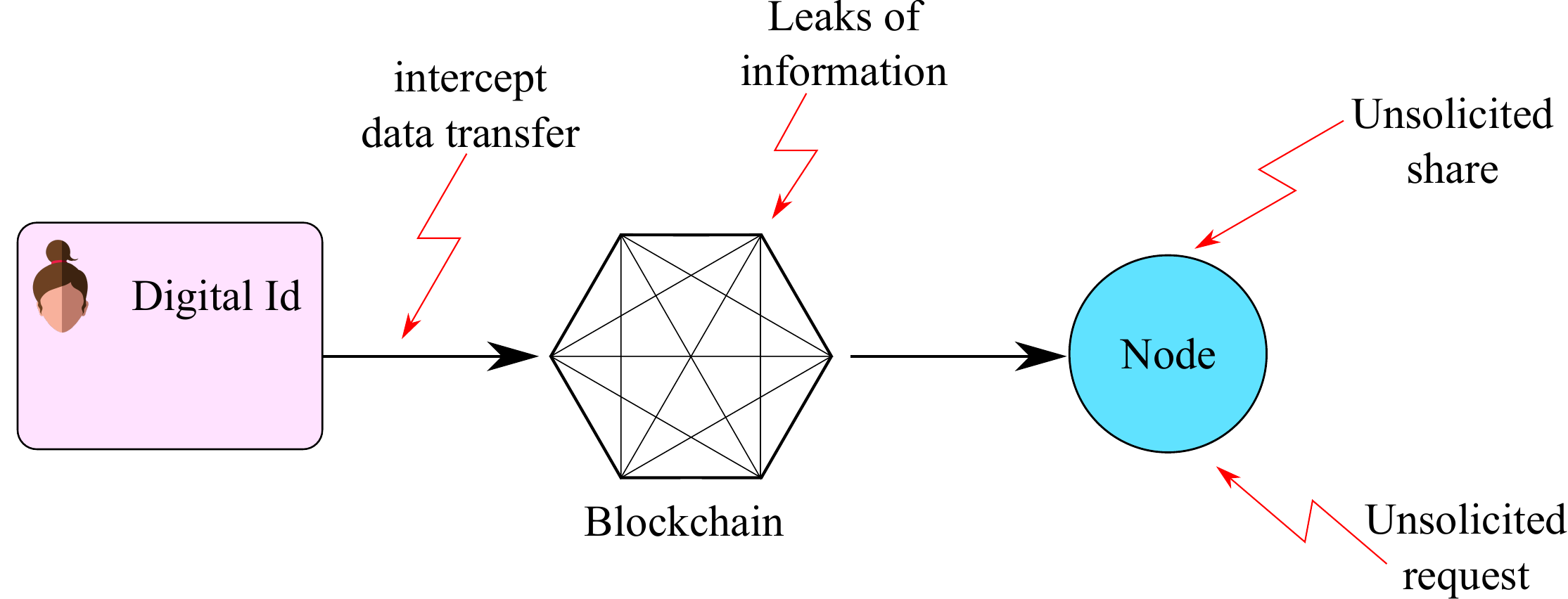}
	\caption{Adversary diagram of the blockchain}
	\label{fig:diagramAdversary}
\end{figure}

The main goal of the BSMD framework is to protect the personal mobility information and secure the privacy of the people in order to fully exploit the benefits of actively or passively solicited large-scale data. To measure the level of protection in BSMD, four groups of adversaries are identified whose attacks can be prevented or hindered thanks to the use of the blockchain. Groups of adversaries may attack the network, the nodes or when the information is been transferred (see Figure~\ref{fig:diagramAdversary}). The identified groups of adversaries are:
\begin{enumerate}
	
	\item \emph{Data interception}: all information transactions are done via a unique and secure \emph{peer-to-peer} connection and information is sent with asymmetric encryption. Attacking a single node may not be worth the effort required to decrypt the data and tamper the connection every time a transactions is made. 
	\item \emph{Data leaks}: all the personal information are decentralized and secured for every individual, so massive leaks on information will require extensive amounts of power to hack the data from a meaningful number of individuals. Also, the interceptions of multiple connections at the same time will require the interception of all connections. The computing power to do this task may not be viable.
	\item \emph{Unsolicited sharing of information}: every node has full access to the ledger so they can easily verify that their information is where they want it to be.   
	\item \emph{Unsolicited request of information}: \emph{smart contracts} let the node decide the information they want to share with specific nodes.
\end{enumerate}
\label{sec:adversaries}

%%%%%%%%%%%%%%%%%%%%%%%%%%%%%%%%%%%%%%%%%%%%%%%%%%%%%%%%%%%%%%%%%%%%%%%%%%%%%%%%%%%%%%%
\section{Conclusions and future work}
\label{sec:Conclusions}

Blockchain framework for Smart Mobility Data-market (BSMD) is designed to solve the privacy, security and management issues related to the sharing of passively as well as actively solicited large-scale data. Data from the individuals, governments, universities and companies are distributed on the network and stored in a decentralized manner, the data transactions are recorded and must have the authorization of the owners. 

To the authors' knowledge there are no references on the minimum components of a blockchain. In the context of transportation, we identified three major components in a generic blockchain, a shared ledger, a \emph{peer-to-peer} network and consensus mechanisms. For the transaction of the mobility information \emph{smart contracts} are added to the BSMD nodes so that they can control the information they want to share. Based on the components of blockchain and the elements involved in the mobility data solicitation we developed a six layer model for BSMD. Theses layers contain the elements necessary to privately and securely transact mobility information. 

We developed three set of simulated demonstrations for the proposed BSMD. In the first experiment we implemented BSMD on a local machine with i7-8cores at 3.5GHz and 16GB RAM. 6 \emph{active} and 100 to 200 \emph{passive} nodes were simulated. \emph{Active} nodes were part of the consensus and ledger management process and would represent entities like municipalities, universities, and trusted private firms. \emph{Passive} nodes were sharing real-time mobility information over a 24hr period. They represented individual smartphones and mobile devices. In the second experiment 10 \emph{active} and 360 \emph{passive} BSMD nodes were simulated. Nodes were running on heterogeneous and geographically-separated devices communicating on a physical network. All and all, for low to medium loads the latency was less than a second and throughput was higher than 99\%. The worst case latency of 6.59sec and throughput of 90\% was observed for heavy loads. In real-life applications, the system performance can further be improved by employing some high performance computing nodes that can do the heavy-lifting, agreeing upon bandwidth consumption strategies, and using distributed ledger technologies e.g. IOTA. In the third set of experiments we demonstrated how BSMD impacts the cybersecurity and privacy of the individual users by securing them against spoofing and message interception as well as giving them control in terms of revoking access to their information using smart contract.

%{In the case study, we demonstrated the working of BSMD marketplace on smartphone based mobility data app developed for Markham, Canada. We also showed that for a large number of active nodes (50 pairs) with low processing power (1GHz) that are simultaneously communicating (100 transactions per node), the system runs with minimal latency i.e. all the transactions are completed in 55 seconds. In the case of communication between only two nodes, the time linearly increased by 3.2 seconds for every 1000 additional transactions. The system performance can further be improved by employing some high performance computing nodes that can do the heavy-lifting, agreeing upon bandwidth consumption strategies, and using distributed ledger technologies e.g. IOTA.}

One possible line of future research is the use of differential privacy based privacy aware machine learning, where the model using the data for parameter estimation preserves the privacy of individuals. Furthermore, instead of the current design in which service provider nodes shares the data with requesting nodes, an alternative design is followed in which the requesting node sends the services provider nodes the analysis model and parameters to be estimated. The service provider nodes run the models on their data and return the analysis parameters back to the requesting node. For the computational time and the model value, the service provider nodes may charge certain fees. 

There are many current real-life cases where BSMD can be implemented. From the perspective of the user, smartcard has the advantage that with a single card it is possible to pay for the all or most of the transit modes. For instance, in Montr\'eal one can use OPUS card for the metro system, regional rail, and local bikeshare. Governments use the entry and exit point logs of the Smartcard  to create strategic plans for the city. Since in many parts of the world the transit is managed by different companies or authorities, if each card reader is connected to BSMD, all payments will be transferred directly to an operator with no third parties involved. Furthermore, each transaction can be assigned to a different DID, so it will be difficult to correlate all the trips of a single user. Transit providers can increase efficiency by eliminating the intermediaries for managing payments while the data of users would still be private and secure.

Another potential beneficiary of the BSMD is the Connected and Autonomous Vehicle (CAV). When the market penetration rate of CAV starts to rise, the BSDM could be used as the backbone network for vehicle-to-vehicle and vehicle-to-infrastructure communication. A secure and resilient network is important as it can be used for implementing intelligent traffic management as well as distributed optimization of the CAV routes \cite{Djavadian2018}. In the CAV network the current location, origins, and destinations of the users are known. So, it is necessary that all communications are secured as well as the individual privacy is preserved.

The analysis and processing of personal mobility data can improve our transportation systems and make our lives more comfortable whether in terms of going to work, building new facilities or reducing the carbon footprint. However, personal mobility data includes several aspects of life that must be private, and if researchers, governments, and companies want to use personal data they must respect the basic human right of privacy. BSMD is designed and built to give people the control over their information and the right to privately share their information. 

\section*{Acknowledgement}
We greatly appreciate Nikki Hera-Farooq's time and efforts in helping us improve the quality of this manuscript. 

\bibliographystyle{abbrv}
\bibliography{bibliography.bib}

\appendix

\section{Incentive Game on BSMD}
\label{sec:appedixgame}
{Assume that a company may or may not give monetary rewards to the users for them sharing mobility information. And a user may or may not share their personal mobility information. Users decide to share information after observing the decisions of the company. A dynamic game is used to model this problem, as in such games the players are able to observe other player's moves before making a decision.}

{Let $G$ be a dynamic game with a set of rational players $N = \{u,c\}$, where $u$ represents a individual user and $c$ represents a company. Let $A = \{A_u, A_c\}$ be the set of actions in the game where $A_u = \{\text{share},\text{not share}\}$ correspond to the actions of the player $u$ and $A_c = \{\text{rewards},\text{no rewards}\}$ corresponds to the actions of player $c$. Each player has associated variables to model the utility of the actions. Player $u$ variables are:}
\begin{itemize}
	
	\item $r_n$: non-monetary reward for sharing information. This reward is the personal value of obtaining aggregated information like routes, waiting times, service hours, etc.
	\item $r_m$: monetary rewards for sharing information.
	\item $c_d$: direct cost of sharing information. Monetary costs of sharing information, like the mobile-data bandwidth usage.
	\item $c_i$: indirect cost of sharing information. The personal value of disclosing private data, higher battery usage, possible decrease in the mobile hardware performance for other tasks, etc.    
\end{itemize}
{Player $c$ variables are:}
\begin{itemize}
	
	\item $c_r$: cost of giving rewards.
	\item $c_f$: cost of collecting fake data. For companies collecting huge amounts of fake data will affect their profit as they cannot sell this data. 
	\item $B$: profits of having/selling personal mobility data
	\item $D$: demand of personal mobility data
\end{itemize}

\begin{figure}[!h]
	\centering
	\includegraphics[width=0.50\textwidth]{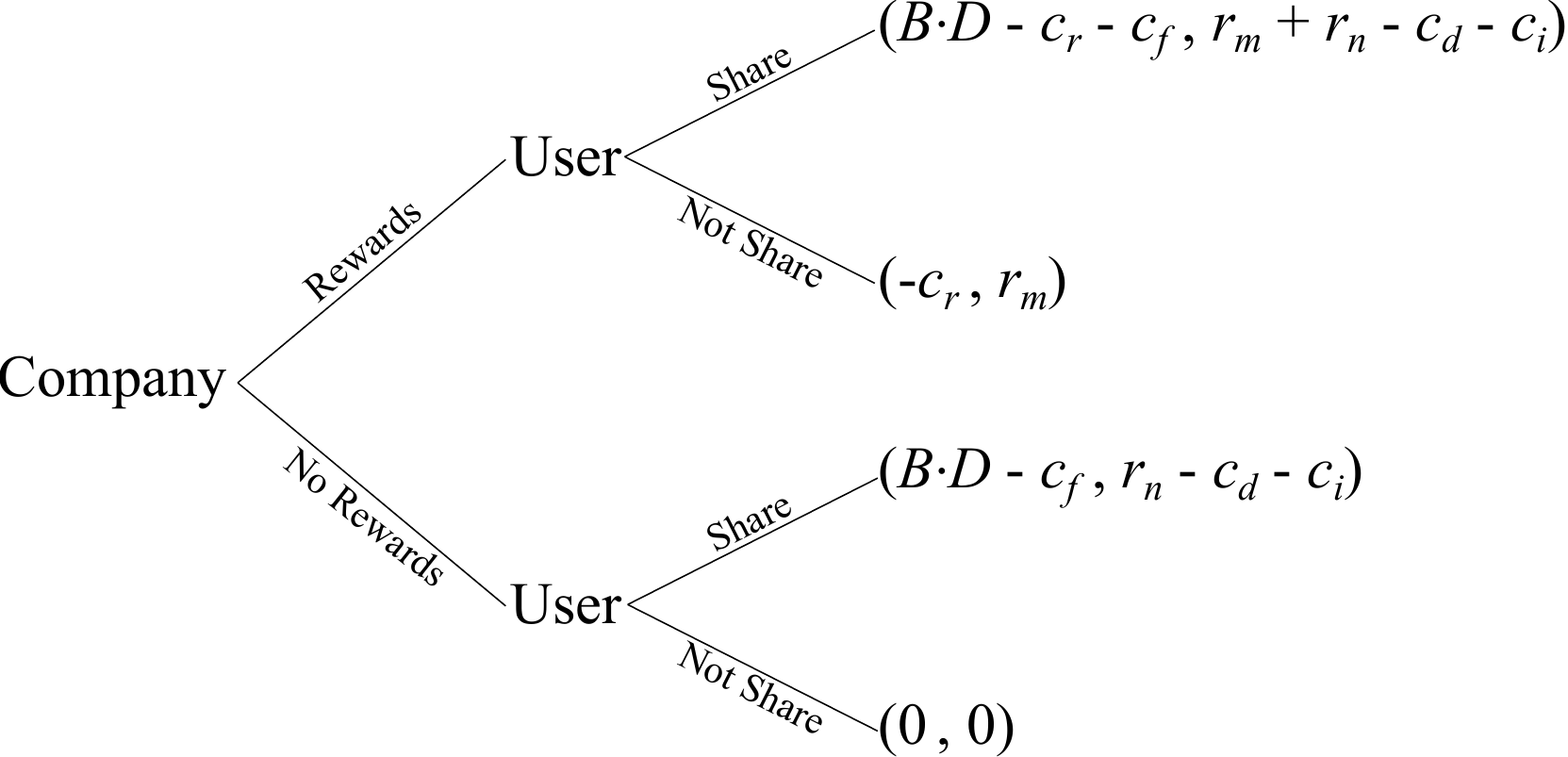}
	\caption{Game three mobility-sharing dynamics}
	\label{fig:game_model}
\end{figure}

{Figure~\ref{fig:game_model} shows the tree structure of $G$. At first the company decides whether to reward or not the users for sharing personal mobility information. After which, users may decide to share or not their personal mobility information with the company. The utilities of the company and user are placed at the end of each branch. For example, in branch \emph{rewards/share}, the utility of the company is $B\cdot D - c_r - c_f$, whereas the utility of the users is $r_m + r_n - c_d - c_i$}.

{By using backward induction, the equilibrium of $G$ can be found. First assume that users will share information regardless of the company giving them monetary rewards or not. Hence, the user considers their utilities are positive under \emph{rewards} and \emph{no rewards}, i.e., $r_m + r_n - c_d - c_i > 0$ and $r_n - c_d - c_i > 0$, both equations can be simplified in Equation~\ref{eq:user_shares}}:

\begin{equation}
	r_n > c_d + c_i \label{eq:user_shares}
\end{equation}

{Given that the user will always share information, the best move for the company would be: \emph{no rewards}. Since $B\cdot D - c_f$ is always greater than $B\cdot D - c_r - c_f$. Hence if Equation~\ref{eq:user_shares} holds the branch \emph{no rewards/share} is in equilibrium. For the Equation~\ref{eq:user_shares} to be true, it means that the user considers the non-monetary rewards for sharing information (the personal value of obtaining aggregated information) to be more valuable than the direct and indirect cost of sharing personal information.}

{To illustrate how the above case will work on BSMD, consider the following example. Suppose that a company does not want to give monetary rewards for personal mobility data and they will only provide a service in exchange for the data. If a user requires the services of the company and they decide to share their information, this will mean that they consider: (a) the information they are sharing does not affect his privacy or (b) they protect their privacy by hiding their real location with \emph{GeoInd}. In this case for the user $c_d + c_i$ is small enough so Equation~\ref{eq:user_shares} holds.} 

{\emph{Smart contracts} are used to automatically enforce the actions of both players so the outcome of the game reach an equilibrium state, which means that both players consider the transaction fair. In this case the \emph{smart contract} terms of the company are: (1) get mobility data, (2) provide a service and (3) no monetary rewards. In order to preserve an equilibrium state in the game, given that the company will not give monetary rewards, the \emph{smart contract} terms of the user are: (1) share the data using \emph{GeoInd}, (2) get a service, and (3) no monetary rewards. If both parties agrees to the terms of the \emph{smart contract} the transaction is performed and both players receive what they want.}

{Now suppose that the users will share information only if they receive a reward. In this case the users utility under the branch \emph{reward/share} is positive while the utility under the branch \emph{no reward/share} is negative. Hence in this scenario Equations~\ref{eq:3} and~\ref{eq:4} hold:}

\begin{align}
	r_m + r_n - c_d - c_i &> 0 \label{eq:3}\\
	r_n - c_d - c_i &< 0 \label{eq:4}
\end{align} 

{Here, the users will share only if they receive rewards, hence the company will choose \emph{rewards} if the utility of the branch \emph{rewards/share} is greater than the utility of the branch \emph{No reward/no share}, i.e., $B\cdot D - c_r - c_f > 0$. Suppose that the company will give a reward to the user for sharing mobility information, hence the value and demand of the information must be considerably larger than the costs associated with collecting the data. Valuable mobility information often require some level of geographic accuracy and additional personal information e.g. gender and/or age range that can be used to develop profile models of people moving in certain areas. Such profile models have high value in the market since they can be useful for strategic marketing or government can use them for mobility studies.}

{The user will share private data, but the costs associated of sharing personal data are considerably larger than the non-monetary rewards. Hence, the user needs extra motivation (monetary rewards) so that their utilities become positive. Note that the company here may choose not to provide a service in exchange for the data since users are more interested in monetary rewards to make their utilities positive.}

{Like in the \emph{no reward} case, \emph{smart contracts} are used to preserve the game equilibrium when the company chooses to give rewards. For instance, the terms of the company can be: (1) get mobility information, age range and gender, and (2) $y$ amount of rewards for the information. While the terms of the user be: (1) receive $x$ amount of rewards for his mobility information, age range and gender. If $x \leq y$ both parties can approve the \emph{smart contract} and perform the transaction.}

{It is clear that if the user decides not share information the company should not give rewards. Hence in $G$, the branch \emph{no rewards/no share} is in equilibrium. Lastly, the branch \emph{rewards/no share} is not in equilibrium. It is dominated by the branch \emph{no rewards/no share} because by definition $c_r$ is a positive value. Both cases, \emph{no rewards/no share} and \emph{rewards/no share}, mean that either the user or company does not agree to the terms. The \emph{smart contract} will enforce the user and company to reach an equilibrium state by stopping the transaction and making the utilities of both players equal to zero.}

\section{Scalability Analysis of BSMD}
\label{sec:appedix}
\subsection{Real-time Transactions}
\begin{figure}[h!]
	\centering
	\includegraphics[scale = .50]{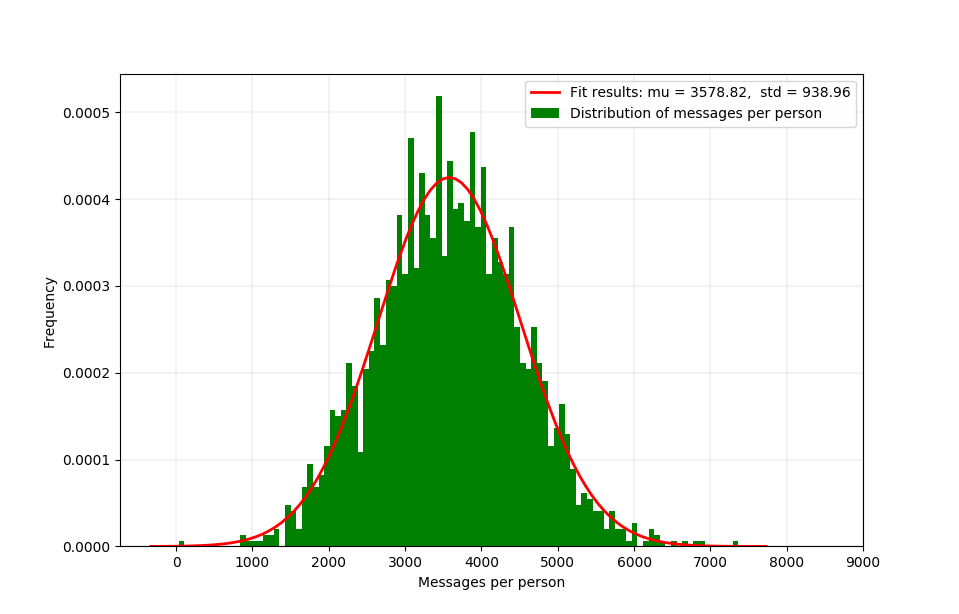}
	\caption{Distribution of real-time location sharing for a given population}
	\label{fig:distAppendix}
\end{figure}

Using Hyperledger-Fabric, BSMD can handle $3,500$ transactions per second with $100\%$ throughput~\cite{Androulaki2018}. For avoiding congestion on BSMD, we can consider strategies that prioritize the transactions nodes send to BSMD. 
%Since transactions of regular travellers sharing their locations in real-time will take most of the traffic, we can compute the maximum number of users BSMD can handle, assuming that the throughput of the network is over $90\%$.
%If an individual sends location points every 5sec, the maximum number of locations they could share in 24hrs is $17,280$, while the minimum is $0$. 
Location based services (LBS) can be used in a broad range of services, but tracking services like real-time traffic alerts or turn-by-turn navigation are the ones that consume most of the transactions, as tracking services need to know the position of the user every few seconds. 
On a typical work day, for navigation and routing assistance, individuals will use LBS for home-work trips and errands/entertainment. If the daily total travel time of an average individual is $5$ hours ($3$hrs for home-work-home trips and $2$hr for errands/entertainment), then an average individual will share approximately $3,600$ GPS points throughout the day.
For a population of $n$ individuals, it is assumed that the number of real-time locations generated by individuals is normally distributed with mean, $\mu = 3,600$, and standard deviation, $\sigma = 950$ as shown in Figure~\ref{fig:distAppendix}.

\begin{figure}[h!]
	\centering
	\includegraphics[scale = .50]{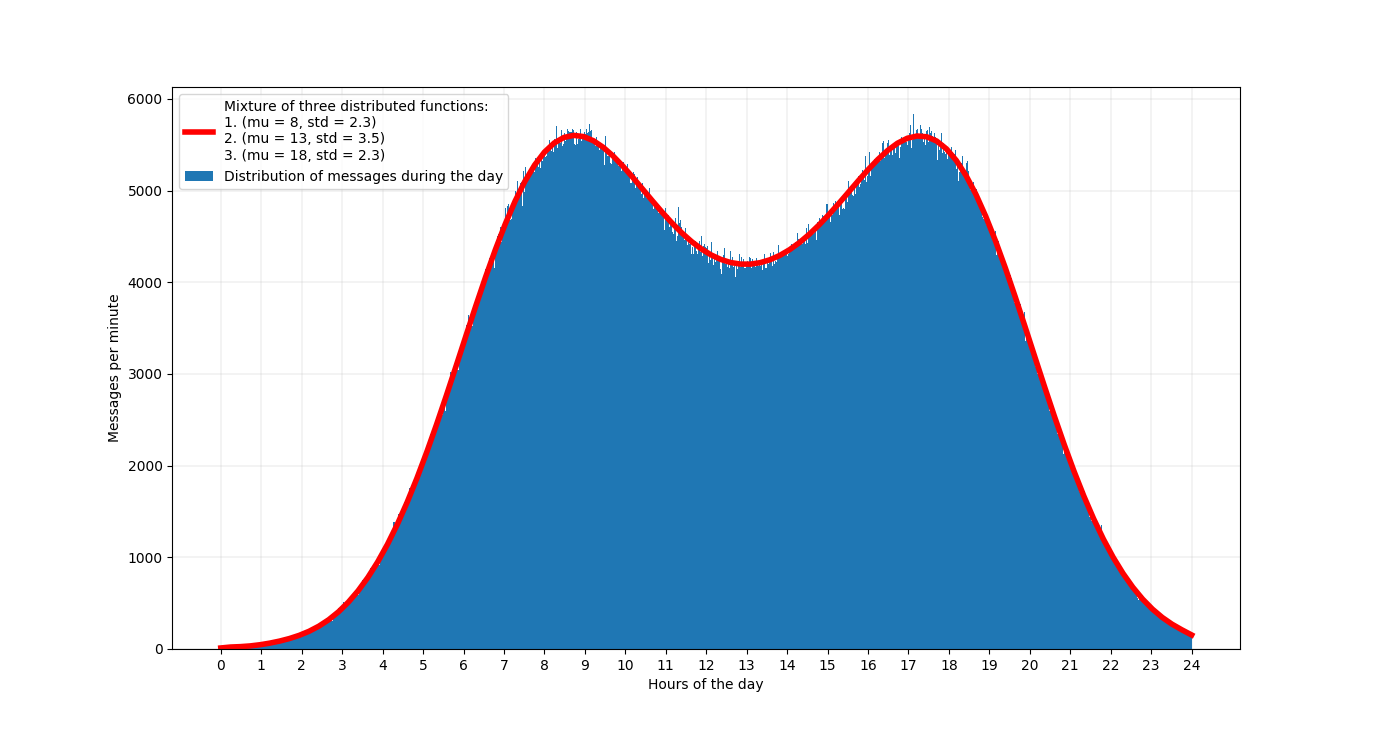}
	\caption{Distribution of messages during the day}
	\label{fig:distAppendix24}
\end{figure}

We can also assume that most trips are around 08:00 (morning peak) when people go to work and around 18:00 (afternoon peak) when people return to home. The distribution of the messages a population sends during the day is shown in Figure~\ref{fig:distAppendix24} and is simulated with a mixture of three normal distributions, such that $\mu_1 = 8$ and $\sigma_1 = 2.3$ represent home-work trips, $\mu_2 = 13$ and $\sigma_2 = 3.5$ represent lunch trips and $\mu_3 = 18$ and $\sigma_3 = 2.3$ represent work-home trips.

By considering the distributions shown in Figures~\ref{fig:distAppendix} and~\ref{fig:distAppendix24}, BSMD will handle $48,000$ users with a steady throughput of $100\%$, while if it is acceptable that an average throughput of $90\%$ during the peak hours (roughly $3,800$ transaction per second) is maintained, BSMD could provide service to $60,000$ users. Therefore, if while travelling everyone is sharing their location in real-time, BSMD could provide a good service to large towns.

Though, in reality not all mobility data are needed to be collected in real-time. It is possible to prioritize which information need to be send in real-time and which information can wait. To provide quality information LBS needs real-time transactions, but people who are comfortable with their normal commute may not always require navigation to get to work/home. These type of individuals may only opt to sell their data as it has a value for the companies. In this case it is proposed that the individuals who do not use LBS for directions, but still want to sell their data can store their mobility patterns in their device and send the information in off-peaks when the traffic on BSMD is low. Table~\ref{tab:max_users} shows the maximum number of users the BSMD can handle if $100\%$, $75\%$, $50\%$ and $25\%$ of travelers use LBS and the BSMD is at $90\%$ throughput during the peak hours. 

\begin{table}[h!]
	\centering
	\caption{Maximum number of users in the BSMD given the percentage of travelers using LBS}
	\label{tab:max_users}
	\begin{tabu} to \linewidth  {| X[c,m] | X[c,m] | X[c,m] |}
		\hline
		\textbf{Percentage of travellers using LBS} & \textbf{Max. number of users in the BSMD at 90\% throughput during peak hours} & \textbf{Max. number of users in the BSMD at steady 100\% throughput} \\ \hline
		$100\%$     & $60,000$    & $48,000$    \\ \hline
		$75\%$      & $76,000$    & $72,000$    \\ \hline
		$50\%$      & $115,000$   & $108,000$   \\ \hline
		$25\%$      & $230,000$   & $216,000$   \\ \hline
	\end{tabu}
\end{table}

\subsection{Computing and Storage Requirements}
\label{sec:appedix_Computing}

{Since BSMD is \textit{public closed} the consensus algorithms require less computing power and can handle more transactions per second. 
	%Taking into consideration the 3500 transaction cap on \emph{Hyperledger-Fabric}, we will discuss the computing need to run it on an \emph{active} node. 
	In an experiment done by \cite{Androulaki2018}, Hyperledger-Fabric reached 3500 transactions with 4 virtual machines of 4, 8, 16 and 32CPU each. The block size was limited to 2MB containing at in average 670 transactions of 4.33kB each. To reach the 3500 transaction, it is expected that some of the nodes will run on a 32CPU machine, while these computers are expensive for personal use (approximately \$4,000), for companies and government the price is manageable. Still, some nodes can have computers with 16 or lesser CPUs that are around \$1,500. Thus the hardware cost is not an impediment to became an \emph{active} node---in terms of computing power in most circumstances, individuals with a 2 year old laptop could host an \emph{active} node.}

{The other aspect that needs to be considered is the storage capacity. \emph{Active} nodes stores full copies of the ledger. Hence they need to consider storage capacity issues. In the experiment by~\cite{Androulaki2018}, the average size of a transaction was 4.33kB. A route from point A to B can be encoded using a 32bit poyline\footnote{\url{https://developers.google.com/maps/documentation/utilities/polylinealgorithm}}. If we do so, a transaction will have 4400 characters left (around 2000 words) for additional information. In reality, it is not necessary to store that much information in a mobility related transaction, where the most important information regarding a trip could be compressed and stored in approximately 100 characters. So, let us assume that the transaction size is around 132Byte where 32Byte are for geographical information and the rest is for the description of the transaction. The header of a block is approximately 80Bytes and a block can contain more than 2,000 transactions of size 132Byte. We also add 2Bytes to each transaction to account for the block related information.}

{If 3,500 transactions are processed every second and the size of each transaction is 134byte the ledger grows 0.46MB every second. Hence, every year the BSMD will grow around 13.8TB a year. Nowadays, standard PCs have around 3-4TB of storage capacity and the storage prices are getting cheaper every year. At this point with a maximum cap of 3,500 transactions/sec it is possible for almost anyone to become an \emph{active} node.}

{Assuming the 3,500 cap could be incremented in near future. Then BSMD can provide service to a bigger population, but \emph{active} nodes will need to have more storage capacity. In Table~\ref{tab:block_size} we show the storage capacity needed for a certain number of transactions and the number of user that will handle the BSMD.}

\begin{table}[h!]
	
	\centering
	\caption{Yearly rate of grow of the BSMD considering transactions per second}
	\label{tab:block_size}
	\begin{tabu} to \linewidth  {| X[c,m] | X[c,m] | X[c,m] | X[c,m] |}
		%\begin{tabular}{|c|c|c|c|}
		\hline
		\textbf{Transactions per second} & \textbf{Transaction size} & \textbf{Rate of growth} & \textbf{ Total Real-time Users} \\ \hline
		3,500                            & 132byte                   & 13.8TB/yr               & 76,000               \\ \hline
		7,000                            & 132byte                   & 27.6TB/yr               & 152,000              \\ \hline
		10,500                           & 132byte                   & 41.4TB/yr               & 228,000              \\ \hline
		14,000                           & 132byte                   & 55.2TB/yr               & 304,000              \\ \hline
		% \end{tabular}
	\end{tabu}
\end{table}

% \begin{figure}[!ht]
%     \centering
%     \begin{subfigure}[b]{0.7\textwidth}
%         \includegraphics[width=\textwidth]{frontEnd.pdf}
%         \caption{Mobile application collects transportation data and shows transit alternatives and trip statistics}
%         \label{fig:mobile}
%     \end{subfigure}
%     ~ %add desired spacing between images, e. g. ~, \quad, \qquad, \hfill etc. 
%       %(or a blank line to force the subfigure onto a new line)
%     \begin{subfigure}[b]{0.9\textwidth}
%         \includegraphics[width=\textwidth]{dashboard.pdf}
%         \caption{Ryerson university can visualize and analyze all transportation data from the users}
%         \label{fig:dashboard}
%     \end{subfigure}
%     \caption{CarbonCount platform}
%     \label{fig:carbonCount}
% \end{figure}

\section{BSMD Blocks}
\label{sec:appedixblocks}
Figure~\ref{fig:block_single} shows the structure of a block private transaction where no private information is written to the ledger. Such block is composed of the following parts:

\begin{figure}
	\centering
	\begin{subfigure}{.4\textwidth}
		\centering
		\includegraphics[scale = .60]{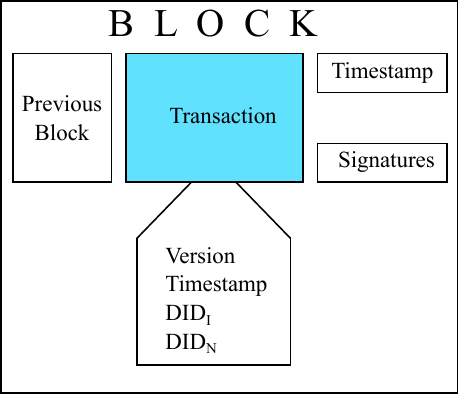}
		\caption{Private data}
		\label{fig:block_single}
	\end{subfigure}
	\begin{subfigure}{.5\textwidth}
		\centering
		\includegraphics[scale = .60]{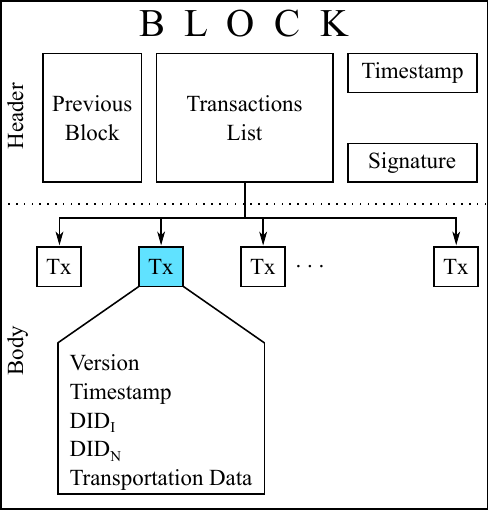}
		\caption{Public data}
		\label{fig:block_iroha}
	\end{subfigure}
	\caption{Block structures in BSMD}
\end{figure}

\begin{itemize}
	
	\item \emph{Previous block}: hash value that chains the block to the previous block.
	\item \emph{Transaction}: dictionary of values relative to the operation. The values are
	\begin{itemize}
		\item \emph{Version}: version number of the blockchain.
		\item \emph{Timestamp}: current time when the transaction was performed.
		\item \emph{DID\textsubscript{N}}: DID of the requester of information.
		\item \emph{DID\textsubscript{I}}: DID of the sender of information.
		\item \emph{Id Broker}: Optional parameter that is used if a \emph{broker} was involved in the transaction.
	\end{itemize}
	\item \emph{Timestamp}: current time when the block was created.
	\item \emph{Signatures}: signatures of \emph{active} nodes, which voted for the block during consensus round. 
\end{itemize} 

{Nodes can keep records of which $DID$s they used for what type of transactions and with whom. Hence, by searching for the $DID$s on BSMD ledger, the nodes can audit the ledger and look for the blocks that contains all the transactions they were part of.}
%In \emph{Indy} blockchain the consensus runs the Redundant Byzantine Fault Tolerance protocol~\cite{Aublin2013}. Other blockchain development kits may use different consensus protocols, but the overall process of the transactions and the formation of blocks is similar.

%{\subsection{Transaction data exploration}}

{For cases where it is desirable to expose the data of transactions, BSMD implemented the identity keys management system. This is done in BSMD by using a combination of \emph{Hyperledger Indy} and \emph{Hyperledger Iroha} functionalities. The blocks in \emph{Indy} contain only the $DID$s related to the nodes performing the transaction. If a third party audits the ledger, it will know that two nodes performed a transaction, but it cannot know the contents of the transaction or the identity of the nodes. While private sharing is desirable for personal information, there exists many cases where the information has to be public for consultation. For example, bike availability at stations is exposed by Bike-share companies so users know in advance the nearest station with available bikes. Having all public information grouped in a single place where anyone can query it will require the usage of multiple APIs in different local servers that can be hacked or tampered.}

{In BSMD, if a company (e.g. Bike-share rental agency) wants to publicly share bicycle availability at stations, each bike-share station can send a transaction to the company node, updating the current state of the station. The complete information of this transaction is written in the ledger so other nodes can query and use the information. The transactions steps here are similar to the case where no information is written. The main difference is in the structure of the blocks and the information that could be written in each block. When a node sends a transaction request to BSMD, it enters to a pool of transactions where it waits to be grouped into a block by an \emph{active} node. Figure~\ref{fig:block_iroha} shows the structure of a block in BSMD for such cases. Such block is composed of the following parts:}

\begin{itemize}
	
	\item \emph{Previous block}: hash value that chains the block to the previous block.
	\item \emph{Transaction list}: array of validated transactions. Each transaction record (Tx) contains the following values:
	\begin{itemize}
		\item \emph{Version}: version number of the blockchain 
		\item \emph{Timestamp}: current time when the transaction was performed
		\item \emph{DID\textsubscript{N}}: DID of the requester of information
		\item \emph{DID\textsubscript{I}}: DID of the sender of information
		\item \emph{Id \emph{Broker}}: Optional parameter that is use if a \emph{broker} was involved in the transaction 
		\item \emph{Transportation data}: exposed transportation data. If the transactions is public the the data is exposed in any other case the data is not exposed
	\end{itemize}
	\item Timestamp: current time when the block was created
	\item Signatures: signatures of \emph{active} nodes, which voted for the block during consensus round
\end{itemize}

\end{document}